\documentclass[pre,twocolumn,showpacs]{revtex4}
\usepackage{epsfig}

\begin{document}

\title{ A Random Walk to a Non-Ergodic Equilibrium Concept}
\author{G. Bel, E. Barkai}
\affiliation{Department of Physics, Bar Ilan University, Ramat-Gan 52900 Israel}
\email{barkaie@mail.biu.ac.il, belgol@mail.biu.ac.il}

\begin{abstract}
{\ Random walk models, such as the trap model, continuous time random walks,
and comb models exhibit weak ergodicity breaking, when the average waiting
time is infinite. The open question is: what statistical mechanical theory
replaces the canonical Boltzmann-Gibbs theory for such systems? In this
manuscript a non-ergodic equilibrium concept is investigated, for a
continuous time random walk model in a potential field. In particular we
show that in the non-ergodic phase the distribution of the occupation time
of the particle on a given lattice point, approaches $U$ or $W$ shaped
distributions related to the arcsin law. We show that when conditions of
detailed balance are applied, these distributions depend on the partition
function of the problem, thus establishing a relation between the
non-ergodic dynamics and canonical statistical mechanics. In the ergodic
phase the distribution function of the occupation times approaches a delta
function centered on the value predicted based on standard Boltzmann--Gibbs
statistics. Relation of our work with single molecule experiments is briefly
discussed. }
\end{abstract}

\maketitle

\section{Introduction}

%

There is growing interest in non-ergodicity of systems whose dynamics is
governed by power law waiting times, is such a way that a state of the
system is occupied with a sojourn time whose average is infinite. Such
non-ergodicity, called weak ergodicity breaking \cite{Werg}, was first
introduced in the context of glassy dynamics. It has found several
applications in Physics: phenomenological models of glassy dynamics \cite%
{Werg}, laser cooling \cite{Bardou}, blinking quantum dots \cite%
{Brokmann,Gennady2}, and models of atomic transport in optical lattice \cite%
{Lutz}. For example single blinking quantum dots, when interacting with a
continuous wave laser field, turn at random times from a bright state in
which many fluorescent photons are emitted, to a dark state. It is found
that the distribution of dark and bright times follows power law behavior.
Somewhat similar statistical behavior is found also for laser cooling of
atoms, where the atom is found in two states in momentum space, a cold
trapped state and a free state, the sojourn time probability density
function has a power law behavior $\psi (\tau )\propto \tau ^{-(1+\alpha )}$
with $\alpha <1$. For such systems the time average of physical observable,
for example the time average of fluorescence intensity of single quantum
dots, is non-identical to the ensemble average even in the long time limit.
From a stochastic point of view such ergodicity breaking is expected, since
the condition to obtain ergodicity is that the measurement time $t$ be much
longer than the microscopical time scale of the problem. However the
microscopic time scale in our examples is infinite, namely the mean trapping
times or the mean dark and bright times diverge. When these characteristic
time scales are infinite, namely $\alpha <1$, we can never make time
averages for long enough times to obtain ergodicity.

It is important to note that the concept of a waiting (i.e. trapping) time
probability density function (PDF) $\psi (\tau )$, with diverging first
moment, is wide spread and found in many fields of Physics \cite%
{Bardou,Bouchaud,Metzler,Havlin,KSZ,Bardou1}. It was introduced into the
theory of transport of charge carriers in disordered material \cite{Scher},
in the context of the continuous time random walk (CTRW). The CTRW describes
a random walk on a lattice with a waiting time PDF of times between jump
events $\psi (\tau )$. The model exhibits anomalous diffusion \cite{Scher}
and aging behaviors \cite{Monthus,Cheng,Cheng1,Grig}, when $\alpha <1$,
which are related to ergodicity breaking. The CTRW found many applications
in the context of chaotic dynamics \cite{Cheng1,Zumofen,ZasRev}, tracer
diffusion in complex flows \cite{swinney,scher}, financial time series \cite%
{Main}, diffusion of bead in a polymer network \cite{Weitz}, to name a few 
\cite{Bouchaud,Metzler}. The dynamics of CTRW is similar to the dynamics of
the comb model \cite{Havlin} and the annealed version of the trap model \cite%
{Monthus}. In turn the trap model is related to the random energy model \cite%
{Arous}. All these systems and models can be at-least suspected of
exhibiting non-ergodic behavior, and hence constructing a general theory of
non-ergodicity for such systems is in our opinion a worthy goal.

Systems and models exhibiting anomalous diffusion, and CTRW behaviors can be
divided into two categories. Systems where the random walk is close to
thermal equilibrium, where the temperature of the system is well defined at
least from an experimental point of view, and non-thermal systems. The
ergodicity breaking of thermal CTRW models, is in conflict with
Boltzmann--Gibbs ergodic assumption. As far as we know, there is no theory
characterizing the non-ergodic properties of the CTRW for either thermal or
non-thermal type of random walk.

Hence one goal of this manuscript is to obtain the non--ergodic properties
of the well known CTRW model on a lattice. Secondly we investigate
ergodicity breaking and its relation to Boltzmann--Gibbs statistics. Using
rather general arguments and using a CTRW model we investigate the
distribution of the total occupation times of a lattice point or a state of
the system. We show that in the limit of long measurement time and in the
ergodic phase the occupation times are obtained using the Boltzmann--Gibbs
canonical ensemble, provided that detailed balance conditions are satisfied.
In the non-ergodic phase we obtain non-trivial distribution of the
occupation times, which are related to the arcsin law. These limiting
distributions are unique in the sense that they do not depend on all the
dynamical details of the underlying model. Further the distributions we
obtain depend on Boltzmann's probability namely on the temperature $T$ and
the partition function $Z$. Thus a connection is established between
non-ergodic dynamics and the basic tool of statistical mechanics.

The study of occupation times in the context of classical Brownian motion
was considered by P. L\'{e}vy. Consider a Brownian path generated with $\dot{%
x}(t)=\eta (t)$, where $\eta (t)$ is Gaussian white noise, in the time
interval $(0,t)$, and with free boundary conditions. The total time $t_{+}$,
the particle spend on the half space $x>0$ is called the occupation time of
the positive half space. The fraction of occupation time $p^{+}=t_{+}/t$ is
distributed according to the celebrated arcsin law \cite{Feller} 
\begin{equation}
\lim_{t\rightarrow \infty }f\left( p^{+}\right) ={\frac{1}{\pi \sqrt{%
p^{+}\left( 1-p^{+}\right) }}},
\end{equation}%
where $0\leq p^{+}\leq 1$. In contrast to naive expectation, it is unlikely
to find $p^{+}=1/2$, which would mean that the particle remains half of the
time in $x>0$. Instead $f\left( p^{+}\right) $ diverges on $p^{+}=0$ and $%
p^{-}=1$, indicating that the Brownian particle tends to stay either in $x>0$
or in $x<0$ for long times of the order of the measurement time $t$. Hence $%
f(p^{+})$ has a $U$ shape. Such a behavior is related to the survival
probability of the Brownian particle. The probability of a Brownian
particle, starting at $x>0$ to remain in $x>0$ without crossing $x=0$,
decays like a power law $t^{-1/2}$. The average time the particle remains in 
$x>0$, before the first crossing of $x=0$, is infinite. Similar $U$ shape
distributions, in far less trivial examples, are investigated more recently
in the context of random walks in random environments \cite{Majumdar},
renewal processes \cite{Godreche}, stochastic processes \cite{Dhar}, zero
temperature Glauber spin dynamics \cite{Dornic}, diffusion equation \cite%
{Newman1}, two dimensional Ising model \cite{Drouffe}, and growing interface 
\cite{Newman}. 

The study of non-ergodicity within the CTRW framework is timely due to
recent single molecule \cite{Jung} type of experiments. In many experiments
anomalous diffusion, and power law behavior was observed using single
particle tracking techniques \cite{Brokmann,Weitz,Virus, Xie,Bizza} (e.g.
single quantum dots \cite{Gennady2}). An interesting example is the
diffusive motion of magnetic beads in an actin network \cite{Weitz}. The
latter exhibit a CTRW type of behavior while the system has a well defined
temperature $T$, namely the random walk seems close to thermal equilibrium
and the particle is coupled to a thermal heat bath. In particular, long
tailed $t^{ - ( 1 + \alpha)}$ waiting time distributions were recorded and
anomalous sub-diffusion $\langle r^2 \rangle \sim t^{\alpha}$ with $\alpha<1$%
, was observed. While clearly ensemble average classification of the
anomalous process, e.g. the mean square displacement, are important, it is
the time averages of single particle trajectories which distinguish the
single particle measurement from standard ensemble average type of
measurement. And stochastic theories of non-ergodicity can help with the
fundamental question in single molecule experiments: are time averages
recorded in such experiment identical to the corresponding ensemble
averages? and if not how do we classify the non-ergodic phase?

This paper is organized as follows. In Sec. \ref{SecLamp} we discuss a
possible generalization of Boltzmann--Gibbs statistics for non-ergodic
dynamics. In Sec. \ref{SecCTRW} we introduce the CTRW model, which yields
the non-ergodic dynamics. Sec. \ref{SecFPT} is the main technical part of
the paper, in which we obtain first passage time properties of the CTRW. The
relation of these properties to the non-ergodic behavior is shown. In Sec. %
\ref{SecMain} we give the main results and compare between the non-ergodic
framework and standard Boltzmann--Gibbs statistics. A brief summary of our
results was published recently \cite{PRL}.

\section{From Boltzmann Statistics to Non-ergodicity}

\label{SecLamp}

In this section we discuss a possible non-ergodic generalization of
Boltzmann--Gibbs theory, without attempting to prove its validity.

The basic tool in statistical mechanics is Boltzmann's probability $P_{x} ^B$
of finding a system in a state with energy $E_x$, 
\begin{equation}
P_{x} ^B = {\frac{\exp\left( - {\frac{E_x }{T}} \right) }{Z}},  \label{EqBol}
\end{equation}
where $T$ is the temperature and $Z= \sum_x \exp( -E_x/T)$. In Eq. (\ref%
{EqBol}) we use the canonical ensemble and assume a classical system, with
discrete energy states $\{0 \le E_1 \le E_2 \cdots \}$. To obtain the
average energy of the system, we use 
\begin{equation}
\langle E \rangle = \sum_x E_x P_{x} ^B,  \label{EqEn}
\end{equation}
and similarly for other physical observables like entropy, free energy etc.
Eq. (\ref{EqEn}) is an ensemble average. When measurement of a single system
is made, a time average of a physical observable is recorded. Consider a
system randomly changing between its energy states $\{ E_x \}$. At a given
time the system occupies one energy state. Let $t_x$ be the total time spent
by the system in energy state $E_x$, within the total observation period $%
(0,t)$. The system may visit state $E_x$ many times during the evolution,
hence $t_x$ is composed in principle from many sojourn times. We define the
occupation fraction 
\begin{equation}
\overline{p}_x = {\frac{ t_x }{t}}, 
\mbox{\ and the time average energy is \
} \overline{E} = \sum_x E_x \overline{p}_x .
\end{equation}
According to statistical mechanics, once the ergodic hypothesis is
satisfied, and within the canonical formalism $\overline{p}_x = P_{x} ^B$
and then $\overline{E} = \langle E \rangle$, and similarly for other
physical observables. More generally, the occupation fraction $\overline{p}%
_x $ is a random variable, whose statistical properties depend on the
underlying dynamics. If Boltzmann's conditions hold the probability density
function (PDF) of $\overline{p}_x $ is 
\begin{equation}
f\left( \overline{p}_x \right)= \delta\left( \overline{p}_x - P_{x} ^B
\right)  \label{Eqdel}
\end{equation}
in the thermodynamic limit. The last Eq. is a restatement of the ergodic
hypothesis. 

In this manuscript we discuss a possible generalization of the ergodic
hypothesis. Our proposal is that the PDF of $\overline{p}_x$, for certain
models described by CTRW type of dynamics, is described by a $%
\delta_{\alpha} $ function 
\[
f\left( \overline{p}_x \right) = \delta_{\alpha}\left( \mathcal{R}_x, 
\overline{p}_x\right)= 
\]
\begin{equation}
{\frac{\sin \alpha \pi }{\pi}} {\frac{ \mathcal{R}_x \overline{p}_x
^{\alpha-1} \left( 1 - \overline{p}_x \right)^{\alpha -1} }{\mathcal{R}_x\
^2 \left( 1 - \overline{p}_x \right)^{2 \alpha} + \overline{p}_x ^{2 \alpha}
+ 2 \mathcal{R}_x \left( 1 - \overline{p}_x \right)^{\alpha}\overline{p}_x
^{\alpha} \cos \pi \alpha}}.  \label{eqGen}
\end{equation}
This PDF was obtained by Lamperti \cite{Lamp} in the context of the
mathematical theory of occupation times (and see Appendix A for details).
For $\mathcal{R}_x=1,\alpha=1/2$ we have the arcsin law. Here we claim that
when local detailed balance condition is satisfied 
\begin{equation}
\mathcal{R}_x= {\frac{P_{x} ^B }{1 -P_{x} ^B}},  \label{eqGen1}
\end{equation}
and $0< \alpha \le 1$. When $\alpha=1$ we get usual ergodic behavior defined
in Eq. (\ref{Eqdel}). Eq. (\ref{eqGen}) is valid only in the limit of long
measurement time. In the non-ergodic phase $\alpha<1$ Eqs. (\ref{eqGen},\ref%
{eqGen1}) establish a relation between the ergodicity breaking and
Boltzmann--Gibbs statistics. The exponent $\alpha$ is the anomalous
diffusion exponent in the relation $\langle x^2 \rangle \propto t^{\alpha}$.

For CTRWs not satisfying detailed balance condition a more general rule
holds. We will show that the PDF of the fraction of time spent on lattice
point $x$, $\overline{p}_x$, is still given by Eq. (\ref{eqGen}). However
now 
\begin{equation}
\mathcal{R}_x= {\frac{P_{x} ^{eq} }{1 -P_{x} ^{eq}}},  \label{eqGen2}
\end{equation}
where $P_{x} ^{eq}$ is the probability that a particle occupies lattice
point $x$ in equilibrium (an equilibrium is obtained for system of finite
size). Here $P_x ^{eq}$ and $P_x ^B$ are probabilities in ensemble sense,
namely if we consider an ensemble of $N$ non interacting particles (or
systems) satisfying some dynamical rule, $P_x ^{eq}$ and $P_x ^B$ yield in
principle the probability that a member of the ensemble occupies state $x$
in equilibrium, which is not identical to $\overline{p}_x$ for non-ergodic
systems.

Let us give some general arguments for the validity of Eqs. (\ref{eqGen},\ref%
{eqGen1}). Consider a particular energy state of the system and call it $%
E_{x}$. At a given time the system is either in energy state $E_{x}$ or is
in any of the other energy states. When the system does not occupy state $x$
we will say that the system is in state $nx$ (not $x$). Assume that sojourn
times in states $x$ and $nx$ are 
\begin{equation}
\psi_{x} (\tau) \sim {\frac{ A_{x} }{|\Gamma\left(-\alpha \right)| \tau^{ 1
+ \alpha} }} \ \ \ \ \ \psi_{nx} (\tau) \sim {\frac{ A_{nx} }{%
|\Gamma\left(-\alpha \right)| \tau^{ 1 + \alpha} }}  \label{eqLTsss1}
\end{equation}
when $\tau$ is large. Also assume that sojourn times in states $x$ and $nx$
are not correlated. Thus we imagine the system occupying state $x$ then
occupying state $nx$, then again state $x$ etc. The amplitudes $A_{x}$ and $%
A_{nx}$ will generally depend on the particular dynamics of the system. We
show in Appendix A that Eq. (\ref{eqGen}) holds with $\mathcal{R}_x =
A_{x}/A_{nx}$. Generally it seems a hopeless mission to calculate the ratio $%
A_{x}/A_{nx}$ from any microscopical model. However a simple physical
argument yields the ratio $\mathcal{R}_x$. Assume that for an ensemble of
systems Boltzmann--Gibbs statistical mechanics holds. Such an assumption
means that on average we must have 
\begin{equation}
\langle \overline{p}_{x} \rangle = P_x ^B,  \label{eqadd1}
\end{equation}
where $P_x ^B$ is Boltzmann's probability of finding a member of an ensemble
of systems in state $x$. On the other hand, Eq. (\ref{eqGen}) yields 
\begin{equation}
\langle \overline{p}_{x} \rangle = \int_0 ^1 \overline{p}_{x} f( \overline{p}%
_{x} ) \mathrm{d} \overline{p}_{x}= {\frac{ \mathcal{R}_x }{1 + \mathcal{R}%
_x }}.  \label{eqadd7}
\end{equation}
Using Eq. (\ref{eqadd1}, \ref{eqadd7}) we obtain Eq. (\ref{eqGen1}).

Our work, is related to the concept of weak ergodicity breaking, suggested
by Bouchaud \cite{Werg}. In standard statistical mechanics, one divides the
phase space of the system, into equally sized cells, and the system is
supposed to visit these cells, with equal probability under certain
constrains (e.g. the energy of the system is constant for the
micro-canonical ensemble). Strong ergodicity breaking means that in order to
leave one phase space cell to another, one has to cross a barrier (e.g. an
energy barrier) which becomes infinite, in the thermodynamic limit. In this
case the time it takes for the system to move from one state to the other is
infinite. It is worth while thinking of such a process, in terms of a
distribution of escape times, $\psi (\tau )=R\exp (-R\tau )$, where $Rt$ is
small and $t$ is the measurement time (e.g. an activation over a very high
energy barrier). In that case the particle/system simply remains in a
certain domain of phase space, for the whole period of observation, and the
system does not explore its entire phase space available for ergodic
systems. A very different scenario was suggested by Bouchaud, in the context
of glassy dynamics and the trap model. If the distribution of sticking
times, follows power law behavior, the average escape time diverges 
\begin{equation}
\langle \tau \rangle \propto \int_{0}^{\infty }\tau \tau ^{-1-\alpha }%
\mathrm{d}\tau =\infty 
\end{equation}%
when $\alpha <1$. Note that also for the strong non-ergodicity case we may
have an infinite waiting time $\langle \tau \rangle =1/R$ when $R\rightarrow
0$. However for power law waiting times the system or particle may still
explore its phase space. Or in other words, exponential waiting times and
power law waiting times, yield very different type of dynamics, even if for
both the average waiting time is infinite. Thus, roughly speaking, for weak
non-ergodicity and for ensemble of particles we may still get
Boltzmann--Gibbs statistics, since from any initial condition the phase
space is totally covered. However the system remains weakly non-ergodic,
since during its evolution, the system will randomly pick one state, which
it will occupy for a very long period (but it still visits all the other
states) and then time averages are not equal to ensemble averages. The goal
of this manuscript is to show that the strong assumptions we used are
correct within a specific model, the well known CTRW model.


\section{CTRW in Force Field}

\label{SecCTRW}

We consider a one dimensional CTRW walk on a lattice. The lattice points are
labeled with index $x$ and $x=-L,-L+1,...,0,...L$, hence the system size is $%
2L+1$. On each lattice point we define a probability $0<Q_{R}(x)<1$ for
jumping right, and a probability for jumping left $Q_{L}(x)=1-Q_{R}(x)$. Let 
$\psi (\tau )$ be the PDF of waiting times at the sites, this PDF does not
depend on the position of the particle. If the particle starts at site $x=0$%
, it will wait there for a period $\tau _{1}$ determined from $\psi (\tau )$%
, it will then jump with probability $Q_{L}(0)$ to the left, and with
probability $Q_{R}(0)$ to the right. After the jump, say to lattice point $1$%
, the particle will pause for a period $\tau _{2}$, whose statistical
properties are determined by $\psi (\tau )$. It will then jump either back
to point $0$ or to $x=2$, according to the probability law $Q_{R}(1)$. Then
the process is renewed. We consider reflecting boundary conditions, namely $%
Q_{L}(L)=Q_{R}(-L)=1$.

The case of a long tailed waiting time distribution, where $\psi(\tau)
\propto \tau^{ - ( 1 + \alpha)}$ when $\tau \to \infty$ and $0<\alpha<1$
yields a non-ergodic behavior. In this case the average waiting time is
infinite. The Laplace transform of $\psi(\tau)$ is 
\begin{equation}
\hat{\psi}\left( u \right) = \int_0 ^{\infty} e^{ - u \tau} \psi\left( \tau
\right) \mathrm{d } \tau.
\end{equation}
As-usual according to Tauberian theorem \cite{Feller}, the small $u$
behavior is 
\begin{equation}
\hat{\psi}(u) \sim 1 - A u^{\alpha} + \cdots  \label{eqususual}
\end{equation}
and $A>0$ is a constant.

Choose a specific lattice point $x$, then define $\theta _{x}(t)=1$ if the
particle is on $x$, otherwise it is zero. We define the occupation fraction
as the time average of $\theta _{x}(t)$, 
\begin{equation}
\overline{p}_{x}={\frac{\int_{0}^{t}\theta _{x}(t^{\prime })\mathrm{d}%
t^{\prime }}{t}},
\end{equation}%
namely $\overline{p}_{x}=t_{x}/t$ where $t_{x}$ is the total time spent on
lattice point $x$ (i.e., the occupation time of site $x$ ). We will later
calculate the PDF of $\overline{p}_{x}$.

Two special cases are the unbiased CTRW, where $Q_L(x) =Q_{R}(x)={1/2}$, and
the biased CTRW with $Q_L(x)=q$. In these cases all transition probabilities
do not depend on the position of the random walker $x$, besides on the
boundaries of-course. In the language of random walks these cases describe
symmetric diffusion process, and diffusion with a drift. Note that in our
model $Q_L(x)$ are not random variables, rather they are included in the
model to mimic a deterministic potential field acting on the system. For
detailed discussion of CTRW models see \cite{Bouchaud,Metzler}.

The case of diffusion with a constant drift, i.e., $q\neq 1/2$ is used many
times to model diffusion under the influence of a constant external driving
force $\mathcal{F}$. If the Physical process is close to thermal equilibrium
the condition of detailed balance is imposed on the dynamics, in order that
for an ensemble of particles Boltzmann equilibrium is reached [see further
discussion after Eq. (\ref{eqDB1})]. The potential energy at each point $x$,
due to the interaction with the external driving force is $E(x)=-\mathcal{F}%
ax$ and $a$ is the lattice spacing. The condition of detailed balance then
reads 
\begin{equation}
{\frac{Q_{L}(x)}{Q_{R}(x)}}=\exp \left( -{\frac{\mathcal{F}a}{T}}\right) 
\label{eqDB}
\end{equation}%
where $T$ is the temperature, and the right hand side of Eq. (\ref{eqDB}) is
independent of lattice coordinate. Since $Q_{L}(x)=q$ is independent of $x$
we have 
\begin{equation}
q={\frac{1}{1+\exp (\mathcal{F}a/T)}}.  \label{Eqq}
\end{equation}

More generally we define an energy profile for the system $%
\{E_{-L},E_{-L+1},...,E_{i},\cdots \}$. The general detailed balance
condition is then 
\begin{equation}
{\frac{Q_{L}(x)}{1-Q_{L}(x-1)}}=\exp \left( -{\frac{E_{x-1}-E_{x}}{T}}%
\right) .  \label{eqDB1}
\end{equation}%
The choice of detailed balance condition means that for an \emph{ensemble}
of particles standard Boltzmann-Gibbs statistics holds. Thus for example if
we observe many independent particles, and look at their density profile in
equilibrium, we will see a profile which is determined by Boltzmann
equilibrium. On the other hand if we consider a trajectory of a single
particle, and from it find $\overline{p}_{x}$ we are not likely to find the
value of $\overline{p}_{x}$ close to Boltzmann's probability, when $\alpha <1
$. Thus ergodicity breaking is found on the level of a single particle. Note
that there is an interesting transition between one particle information and
many particle behavior, however this is not the subject of our work \cite%
{Newman1}.

\section{First Passage Times}

\label{SecFPT}

The problem of ergodicity breaking is related in this section to the problem
of first passage times.

The process $\theta_{x}(t)$ is a two state process, with state $x$ denoting
particle on lattice point $x$ and state $nx$ indicating that the particle is
not on $x$. Obviously the waiting times in state $x$ are given by $%
\psi_{x}(\tau)= \psi(\tau)$. To obtain the PDF of waiting times in state $nx$%
, $\psi_{nx}(\tau)$ we must calculate statistical properties of first
passage times. After the particle leaves point $x$ it is located either on $%
x+1$ or $x-1$ with probabilities $Q_R(x)$ and $Q_L(x)$, respectively. Let $%
t_L$ denote the time it will take the particle to return to $x$ starting at
point $x-1$, i.e. the first passage time from $x-1$ to $x$. Let $t_R$ be the
first passage time to reach $x$ starting from $x+1$. Let $f_R(t_R)$ [$%
f_L(t_L)$] be the PDF of the first passage time $t_R$ [$t_L$] respectively.
Then the PDF of times in state $nx$ is given by 
\begin{equation}
\psi_{nx}\left(\tau\right) = Q_R(x) f_R \left(\tau\right) + Q_L(x) f_L
\left(\tau\right).  \label{eqsum}
\end{equation}
In principle once the long time behavior of the PDFs of first passage times
is obtained, we have $\psi_{nx}\left( \tau \right)$ and $\psi_{x}(\tau)$,
and then we may use the formalism developed in Appendix A to obtain the PDF
of the occupation fraction $\overline{p}_x$. We now investigate the first
passage times PDFs for biased and unbiased CTRW, using an analytical
approach. The reader not interested in mathematical details, may skip to
Sec. \ref{SecMain}.

\subsection{Relation Between Discrete Time and continuous time RWs}

For convenience we define a new lattice. We consider the CTRW in one
dimension, on lattice points $x=0,1,2,...,\tilde{L}$. Point $x=0$ is a
``sticky'' absorbing boundary, namely once the particle reaches point $x=0$
it remains there for ever. Point $\tilde{L}$ is a reflecting boundary, and
initially at time $t=0$ the particle is on $x=1$. Let $S_{\mathrm{CT}}(t)$
be the survival probability of the CTRW particle, and the subscript $\mathrm{%
CT}$ indicates CTRW. The object of interest is the PDF of first passage time 
$f_{\mathrm{CT}}(t)$, which is minus the time derivative of $S_{\mathrm{CT}%
}(t)$. The solution is possible due to an important relation \cite{Weiss}
between the CTRW first passage time problem and that of discrete time random
walks. In \cite{Weiss} first passage time problem with CTRW dynamics with
exponential waiting times was considered.

Point $0$ of the new lattice is point $x$ in the original problem and $%
\tilde{L} = L-x$ and similarly for the other $\tilde{L}-1$ points of the new
lattice. Hence the calculation of the first passage PDF on the new lattice $%
x=0,1,2,...\tilde{L}$ yields $f_R(t_R)$. With straight forward change of
notation we may consider also $f_L(t_L)$.

Let $S_{\mbox{CT}}(t)$ be the survival probability of the CTRW particle in
the interval $x=1,....,x=\tilde{L}$. Let $S_{\mbox{dis}}(N)$ be the
probability of survival after $N$ jumps events, for a particle starting at $%
x=1$, the subscript $\mbox{dis}$ stands for discrete. Then 
\begin{equation}
S_{\mbox{CT}}(t)=\sum_{N=0}^{\infty }S_{\mbox{dis}}(N)P\left( N,t\right)
\label{eqFT01}
\end{equation}%
where $P(N,t)$ is the probability for $N$ steps, in time $t$, in a CTRW
process. In Laplace, $t\rightarrow u$ space it is easy to show using the
convolution theorem of Laplace transform that 
\begin{equation}
\hat{P}(N,u)={\frac{1-\hat{\psi}(u)}{u}}\hat{\psi}^{N}(u)  \label{eqFT02}
\end{equation}%
where $\hat{\psi}(u)$ is the Laplace transform of $\psi (\tau )$. In this
work the discrete Laplace transform of an arbitrary function $G(N)$, also
called the $z$ transform is defined as 
\begin{equation}
\tilde{G}\left( z\right) =\sum_{N=0}^{\infty }z^{N}G\left( N\right) .
\label{eqFT03z}
\end{equation}%
%
%
%
%
%
%
%
Using Eqs. (\ref{eqFT01},\ref{eqFT02}) we find 
\begin{equation}
\hat{S}_{\mbox{CT}}\left( u\right) ={\frac{1-\hat{\psi}(u)}{u}}\tilde{S}_{%
\mbox{dis}}\left[ \hat{\psi}\left( u\right) \right] .  \label{eqFT04}
\end{equation}%
This equation establishes the relation between the discrete and continuous
time problems.

Let $P_{x}(N)$ be the probability of occupying site $x$ after $N$ jumps and $%
P_{x}(0)=\delta _{x1}$. The Master equation describing the discrete time
problem is given by 
\[
P_{0}(N+1)=Q_{L}(1)P_{1}(N)+P_{0}(N)
\]%
since the origin $0$ is absorbing 
\[
P_{1}(N+1)=Q_{L}(2)P_{2}(N)
\]%
\[
P_{2}(N+1)=Q_{R}(1)P_{1}(N)+Q_{L}(3)P_{3}(N)
\]%
\[
P_{x}(N+1)=Q_{R}(x-1)P_{x-1}(N)+Q_{L}(x+1)P_{x+1}(N)
\]%
\[
P_{\tilde{L}-1}(N+1)=P_{\tilde{L}}(N)+Q_{R}(\tilde{L}-2)P_{\tilde{L}-2}(N)
\]%
\begin{equation}
P_{\tilde{L}}(N+1)=Q_{R}(\tilde{L}-1)P_{\tilde{L}-1}(N).  \label{eqFT05}
\end{equation}%
The probability to be absorbed for the first time at $x=0$ after $N+1$ jumps
(the discrete time) is 
\begin{equation}
F_{\mbox{dis}}\left( N+1\right) =Q_{L}(1)P_{1}(N).  \label{eqFT06}
\end{equation}%
The discrete survival probability is given by 
\begin{equation}
S_{\mbox{dis}}(N)=1-P_{0}\left( N\right) .  \label{eqFT07}
\end{equation}%
Using Eq. (\ref{eqFT05}) 
\begin{equation}
S_{\mbox{dis}}(N)=1-\left[ Q_{L}(1)P_{1}(N-1)+P_{0}(N-1)\right] ,
\label{eqFT071}
\end{equation}%
and from Eq. (\ref{eqFT06}) 
\begin{equation}
S_{\mbox{dis}}(N)=1-\left[ F_{\mbox{dis}}\left( N\right) +P_{0}(N-1)\right] .
\end{equation}%
Using Eq. (\ref{eqFT07}) we have 
\begin{equation}
S_{\mbox{dis}}(N)-S_{\mbox{dis}}(N-1)=-F_{\mbox{dis}}\left( N\right) ,
\label{eqFT08}
\end{equation}%
which simply means that the change in the survival probability at step $N$
is equal to minus the probability of first passage. Using the $z$ transform
Eq. (\ref{eqFT03z}) of Eq. (\ref{eqFT08}) we find 
\begin{equation}
\tilde{S}_{\mbox{dis}}\left( z\right) ={\frac{1-\tilde{F}_{\mbox{dis}}\left(
z\right) }{1-z}}.  \label{eqFT09}
\end{equation}%
Hence from Eq. (\ref{eqFT04}) 
\begin{equation}
\hat{S}_{\mbox{CT}}(u)={\frac{1}{u}}\left\{ 1-\tilde{F}_{\mbox{dis}}\left[ 
\hat{\psi}\left( u\right) \right] \right\} .  \label{eqFT10}
\end{equation}%
Let $f_{\mbox{CT}}\left( t\right) $ be the first passage time PDF of the
CTRW problem. As-usual 
\begin{equation}
f_{CT}\left( t\right) =-{\frac{\mathrm{d}}{\mathrm{d}t}}S_{CT}(t).
\label{eqFT11}
\end{equation}%
which is the continuous pair of Eq. (\ref{eqFT08}). If the random walker
always returns to the origin, then $\int_{0}^{\infty }f_{CT}(t)\mathrm{d}t=1$%
, and Eq. (\ref{eqFT11}) yields 
\begin{equation}
\hat{f}_{CT}\left( u\right) =-u\hat{S}_{CT}(u)+1  \label{eqFT12}
\end{equation}%
and using Eq. (\ref{eqFT10}) 
\begin{equation}
\hat{f}_{CT}\left( u\right) =\tilde{F}_{\mbox{dis}}\left[ \hat{\psi}\left(
u\right) \right] .  \label{eqFT13}
\end{equation}%
This is the most important equation of this sub-section. At-least in some
cases the solution of the discrete time first passage time problem, in $z$
space is possible, and then we can transform the solution to Laplace $u$
space of the seemingly more difficult case of continuous time. Note that our
assumption that the random walk is recurrent is valid only when the system
size is finite, and $Q_{L}(x)>0$ for any $x$ besides on the boundary.

\subsection{First Passage Time for Unbiased Case}

We now find the first passage time distribution for the unbiased CTRW in
Laplace space. For the unbiased random walk we have $Q_L(x)=Q_R(x)=1/2$, for 
$x\ne 0$, $x\ne \tilde{L}$. And as mentioned $x=0$ is the absorbing boundary
condition, while $\tilde{L}$ is a reflecting wall. As shown we may consider
the first passage time for the discrete time random walk Eq. (\ref{eqFT05})
and then use the transformation Eq. (\ref{eqFT13}) to obtain the
corresponding CTRW first passage time. Using Eq. (\ref{eqFT03z}) the $z$
transform of Eq. (\ref{eqFT05}) is 
\[
\tilde{P}_0 \left( z \right) = {\frac{ z }{2}} \tilde{P}_1 \left( z \right)
+ z \tilde{P}_0\left( z \right) 
\]
using the initial conditions $P_1(0)=1$, 
\[
\tilde{P}_1 \left( z \right) -1 = {\frac{ z }{2}} \tilde{P}_2 \left( z
\right), 
\]
for $x= 2, \cdots, L-2$, 
\[
\tilde{P}_x \left( z \right) = {\frac{ z }{2}} \left[ \tilde{P}_{x-1} \left(
z \right) + \tilde{P}_{x+1} \left( z \right)\right] , 
\]
\[
\tilde{P}_{\tilde{L}-1} \left( z \right) = z \tilde{P}_{\tilde{L}} \left( z
\right) + {\frac{ z }{2}} \tilde{P}_{\tilde{L} -2} \left( z \right) 
\]
\begin{equation}
\tilde{P}_{\tilde{L}} \left( z \right) = {\frac{ z }{2}} \tilde{P}_{\tilde{L}%
-1} \left( z \right),  \label{eqZTfp01}
\end{equation}
and using Eq. (\ref{eqFT06}) 
\begin{equation}
\tilde{F} _{\mbox{dis}}\left( z \right) = {\frac{ z }{2}} \tilde{P}_1 \left(
z \right).  \label{eqZTfp02}
\end{equation}
To solve these equations we use a recursive solution method \cite%
{Redner,Gold}. We define $\phi_x(z)$ using the relation 
\begin{equation}
\tilde{P}_x \left( z \right) = \phi_x(z) \tilde{P}_{x - 1} \left( z \right),
\label{eqZTfp03}
\end{equation}
and it is easy to show using Eqs. (\ref{eqZTfp01}, \ref{eqZTfp03}) 
\begin{equation}
\phi_{\tilde{L}}(z) = z/2 \ \ \ \phi_{\tilde{L}-1}(z) = (z/2)/ ( 1 - z^2 /2).
\label{eeqZTadd}
\end{equation}
The function $\phi_x (z)$ also satisfies the recursion relation 
\begin{equation}
\phi_{x - 1} \left( z \right) = {\frac{ (z /2 ) }{1 - z \phi_{x} \left( z
\right)/2 }}  \label{eqZTfp04}
\end{equation}
which is easy to obtain from Eq. (\ref{eqZTfp01}). Let 
\begin{equation}
\phi_x \left( z \right) = {\frac{ g_x \left( z \right) }{h_x \left( z
\right) }}  \label{eqZTfp05}
\end{equation}
and using Eq. (\ref{eqZTfp04}) 
\begin{equation}
\left( 
\begin{array}{c}
g_{x - 1}\left( z \right) \\ 
h_{x - 1} \left( z \right)%
\end{array}
\right)= \left( 
\begin{array}{cc}
0 & {\frac{ z }{2}} \\ 
- {\frac{z }{2}} & 1%
\end{array}
\right) \left( 
\begin{array}{c}
g_x \left( z \right) \\ 
h_x \left( z \right)%
\end{array}
\right).  \label{eqZTfp06}
\end{equation}
Since we are interested only in the ratio $g_x \left( z \right) / h_x \left(
z \right) $ we may set $h_{\tilde{L}} \left( z \right) = 1$ and $g_{\tilde{L}%
}(z)=z/2$ using Eq. (\ref{eeqZTadd}). Eq. (\ref{eeqZTadd}) gives the seeds
for the iteration rule Eq. (\ref{eqZTfp06}): $h_{\tilde{L}-1}(z)=1 -z^2/2$
and $g_{\tilde{L}-1}(z)=z/2$, which yield $h_{\tilde{L}-2}(z),g_{\tilde{L}%
-2}(z)$ etc. Let 
\begin{equation}
h_x\left( z \right) = B_{+} \left(\Lambda_{+}\right)^{\tilde{L}-x} + B_{-}
\left(\Lambda_{-}\right)^{\tilde{L}-x}  \label{eqZTfp07}
\end{equation}
and from $h_{\tilde{L}}(z) =1$ we have $B_+ + B_-= 1$. $\Lambda_{\pm}$ are
eigen values of the matrix in Eq. (\ref{eqZTfp06}). 
\begin{equation}
\Lambda_{\pm} = {\frac{ 1 \pm \sqrt{ 1 - z^2} }{2}}.  \label{eqZTfp08}
\end{equation}
Using $h_{\tilde{L}-1}\left( z \right) = 1 - z^2/2$ it is easy to show 
\begin{equation}
B_{-} = {\frac{ 1 - z^2/2 - \Lambda_{+} }{\Lambda_{-} - \Lambda_{+} }}
\label{eqZTfp09}
\end{equation}
and $B_{+} = 1 - B_-$. Using 
\begin{equation}
\tilde{P}_1 \left( z \right) = {\frac{ 1 }{1 - z \phi_2 \left( z \right)/2}}
\label{eqZTfp09a}
\end{equation}
and $\phi_2 (z) = z h_3 (z) / 2 h_2 (z)$ and Eqs. (\ref{eqZTfp02}, \ref%
{eqZTfp07}) we find 
\begin{equation}
\tilde{F}_{\mbox{dis}} \left( z \right) = {\frac{ z/2 }{1 - {\frac{z^2 }{4}} 
{\frac{ B_+ \Lambda_+ ^{\tilde{L}-3} + B_1 \Lambda_- ^{\tilde{L}-3} }{B_+
\Lambda_+ ^{\tilde{L}-2} + \Lambda_{-} ^{\tilde{L}-2} }} }}  \label{eqZTfp10}
\end{equation}
This equation is important since it yields the discrete first passage time
probability with which the CTRW PDF of first passage time can be obtained.
Using the Laplace transform of the waiting time PDF Eq. (\ref{eqususual})
and Eqs. (\ref{eqFT13}, \ref{eqZTfp10}) we obtain the small $u$ behavior 
\begin{equation}
\hat{f}_{\mbox{CT}} \left( u \right) \sim 1 - (2 \tilde{L} -1) A u^{\alpha}
+ \cdots .  \label{eqZTfp11}
\end{equation}
To summarize Eq. (\ref{eqZTfp11}) yields the Laplace transform of the first
passage times of the unbiased CTRW with reflecting boundary condition on $%
\tilde{L}$, absorbing on the origin, and initial location of the particle on 
$x=1$.

\subsection{First Passage Time for Uniform Bias}

We now find the first passage time distribution for the biased CTRW in
Laplace space, skipping many of the algebraic details. Now the probability
to jump left is $Q_L(x)=q$ and hence the probability to jump to the right is 
$Q_R(x)=1-q$, for $x\ne 0$, $x\ne \tilde{L}$. The two boundary conditions
are: $x=0$ is absorbing, while $\tilde{L}$ is a reflecting wall. Like the
unbiased case we treat the problem of the discrete time random walk and then
use the transformation Eq. (\ref{eqFT13}) to obtain the corresponding CTRW
first passage time distribution.

In this case the $z$ transform of the master Eq. (\ref{eqFT05}) is 
\[
\tilde{P}_0\left( z \right) = z q \tilde{P}_1 \left( z \right) + z \tilde{P}%
_0\left( z \right) 
\]
\[
\tilde{P}_1\left( z \right) - 1 = z q \tilde{P}_2 \left( z \right) 
\]
\[
\tilde{P}_x \left( z \right) = z \left( 1 - q \right) \tilde{P}_{x-1} \left(
z \right) + z q \tilde{P}_{x+1} \left( z \right) 
\]
\[
\tilde{P}_{\tilde{L}-1} \left( z \right) = z \tilde{P}_{\tilde{L}}\left( z
\right) + z\left( 1 - q \right) \tilde{P}_{\tilde{L}-2}\left( z \right) 
\]
\begin{equation}
\tilde{P}_{\tilde{L}} \left( z \right) = z \left( 1 - q \right) \tilde{P}_{%
\tilde{L}-1}\left( z \right).  \label{eqBias01}
\end{equation}
And using Eq. (\ref{eqFT06}) 
\begin{equation}
\tilde{F}_{\mbox{dis}} \left( z \right) = z q \tilde{P}_1 \left( z \right).
\label{eqBias02}
\end{equation}
The solution of the biased master equation (\ref{eqBias01}) follows the same
procedure as for the unbiased and yields 
\begin{equation}
\tilde{F} \left( z \right) = {\frac{ q z }{1 - q \left( 1 - q \right) z^2 {%
\frac{ B_{+} \lambda_{+} ^{ \tilde{L} - 3 } + B_{-} \lambda_{-} ^{\tilde{L}
- 3 } }{B_{+} \lambda_{+} ^{ \tilde{L} - 2 } + B_{-} \lambda_{-} ^{\tilde{L}
- 2 } }} }},  \label{eqBias03}
\end{equation}
where 
\begin{equation}
\lambda_{\pm} \left( z \right) = {\frac{ 1 \pm \sqrt{ 1 - 4 q z^2 \left( 1 -
q \right) } }{2}},  \label{eqBias04}
\end{equation}
$B_{+} + B_{-} = 1$, and 
\begin{equation}
B_{+}(z) = {\frac{ 1 - \lambda_{-} - z^2 \left( 1 - q \right) }{\lambda_{+}
- \lambda_{-} }}.  \label{eqBias05}
\end{equation}
Using Eq. (\ref{eqBias02}) one can show that $\tilde{F}_{\mbox{dis}}(z=1) =
1 $, for any finite $\tilde{L}$ and $q\ne 0$, namely if we wait long enough
the particle always reaches the sticky boundary on $x=0$.

We now return to the CTRW problem. We use the relation Eq. (\ref{eqFT13})
and insert in Eq. (\ref{eqBias03}) the small $u$ behavior of the Laplace
transform of the waiting time PDF Eq. (\ref{eqususual}). In the limit of $%
u\rightarrow 0$ we find the Laplace transform of the PDF of the first
passage time of the CTRW particle 
\begin{equation}
\hat{f}_{CT}\left( u\right) \sim 1-{\frac{Au^{\alpha }}{2q-1}}\left[
1-2\left( 1-q\right) \left( {\frac{1-q}{q}}\right) ^{\tilde{L}-1}\right]
+\cdots .  \label{eqBias06}
\end{equation}%
This is the main result of this section, since it will yield the non-ergodic
properties of the biased CTRW. We see that for $q=1$ or $L=1$, $\hat{f}%
_{CT}\left( u\right) \sim 1-Au^{\alpha }$ as expected since then $\hat{f}%
_{CT}(u)=\hat{\psi}(u)$. The second term on the right hand side of Eq. (\ref%
{eqBias06}) will diverge when $q<1/2$ and $L\rightarrow \infty $, as
expected for an infinite system, and for a random walker moving against the
average drift. We see from Eq. (\ref{eqBias06}), that the PDF of first
passage times $f_{\mbox{CT}}(t)\propto t^{-(1+\alpha )}$. in the limit of
long times, when $\alpha <1$. In the limit $q\rightarrow 1/2$ the solution
for the biased case Eq. (\ref{eqBias06}), reduces to the unbiased solution
Eq. (\ref{eqZTfp11})
\begin{figure}[tbp]
\begin{center}
\epsfxsize=70mm \epsfbox{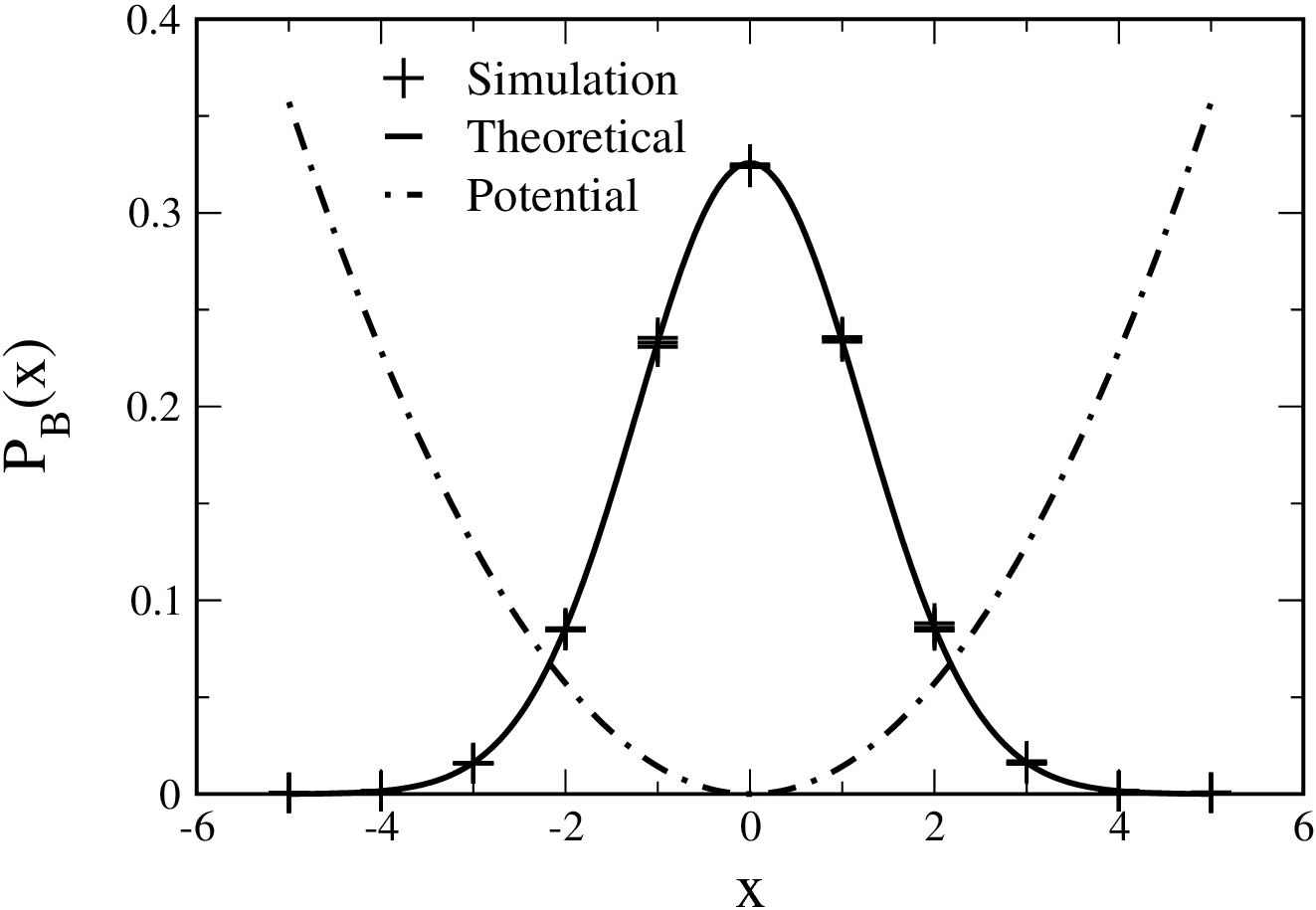}
\end{center}
\caption{ Boltzmann's equilibrium for an ensemble of CTRW particles in an
harmonic potential field, and fixed temperature. In simulations (cross) the
CTRW particle with $\protect\alpha=0.3,0.5$ and $\protect\alpha=0.8$ was
considered. The figure illustrates that for ensemble of particles, standard
equilibrium is obtained, ergodicity breaking is found only when long time
averages of single particle trajectories are analyzed. The scaled potential
(dot dash curve) is the harmonic potential field, and the theoretical curve
is Boltzmann equilibrium distribution. To construct the histogram we used $%
N=10^6$ particles, temperature $T=3$, and the total observation time $t=10^6$%
. }
\label{fig1}
\end{figure}

\section{Main Results}

\label{SecMain}

\subsection{Non- Thermal random walks}

First consider the unbiased one dimensional CTRW on a lattice $x=-L,\cdots,L$%
. The PDF of the fraction of occupation time $\overline{p}_{x}=t_{x}/t$ on a
lattice point $x$, excluding the boundary points, is obtained using Eqs. ( %
\ref{eqsum}, \ref{eqZTfp11}, \ref{eqAm25}, \ref{eqGA11} ). The general idea
of the proof is to note that $\hat{\psi}_x(u) =\hat{\psi}(u)$ 
\begin{equation}
\psi_{nx}(u) \sim 1 - A \left( 2 L - 1 \right) u^{\alpha}
\end{equation}
for $u \to 0$ and hence using Appendix A we find 
\begin{equation}
\lim_{t \to \infty} f \left( \overline{p}_{x} \right) = \delta_{\alpha}
\left((2 L -1)^{-1},\overline{p}_{x}\right).  \label{eqGMR00}
\end{equation}
Where the $\delta_{\alpha}$ function was defined in Eq. (\ref{eqGen}). Eq. (%
\ref{eqGMR00}) does not depend on the position $x$ of the observation point,
reflecting the symmetry of the problem. From Eq. (\ref{eqGMR00}) we see that
the amplitude ratio satisfies $\mathcal{R}_x=1/(2 L -1) < 1$ when $L>1$.
This inequality means that we are less likely to find the particle on the
particular lattice point $x$ under observation (state $x$), if compared with
the probability of finding the particle on any of the other lattice points
(state $nx$).

For the biased random walk, when the probability of jumping left is $q$, we
consider the PDF of fraction of time $\overline{p}_{x}$ on a lattice point $%
x $. Now clearly different locations have different distributions of the
fraction of occupation time, reflecting the fact the the system is biased.
The Laplace transform of the sojourn times on $x$ is simply 
\begin{equation}
\hat{\psi}_{x}\left( u \right) = \hat{\psi} \left( u \right) \sim 1 - A
u^{\alpha}
\end{equation}
and using Eq. (\ref{eqsum}) the sojourn times in all other states (nx) is 
\begin{equation}
\hat{\psi}_{nx} \left( u \right) = \left( 1 - q \right) \hat{f}_R \left( L -
x, u\right) + q f_L \left( x + L , u \right).  \label{sum2}
\end{equation}
Here $\hat{f}_R \left( L - x, u\right)$ is the Laplace transform of the
first passage time PDF, for a system of size $L-x + 1$, obtained in Eq. (\ref%
{eqBias06}). Similarly for $\hat{f}_L \left( L - x, u\right)$ however now
replace $q$ with $1-q$ in Eq. (\ref{eqBias06}). Using Eq. (\ref{sum2}) we
find the small $u$ behavior of $\hat{\psi}_{nx}(u)$, and then using Eq. (\ref%
{eqAm25}) we find 
\begin{equation}
\hat{\psi}_{nx}(u) \sim 1 - {\frac{A }{\mathcal{R}_x}} u^{\alpha} + \cdots
\label{eqcoco}
\end{equation}
\begin{widetext}
\begin{equation}
 {\cal R}_x  = 
\left\{ {2 \over 2 q - 1} \left[ q^2 \left( {q \over 1 - q } \right)^{L + x -1} - 
\left( 1 - q \right)^2 \left( { 1 - q \over q} \right)^{L-x-1}\right] - 1\right\}^{-1}.
\label{eqGMR03}
\end{equation}
The latter Eqs. (\ref{eqcoco},\ref{eqGMR03}) and the results obtained
in Appendix A indicate that the PDF of fraction
of occupation time is 
\begin{equation}
f \left( \overline{p}_{x} \right) =\delta_{\alpha}
\left( {\cal R}_x, \overline{p}_{x} \right) 
\label{eqGMRo2}
\end{equation}
with ${\cal R}_x$ given in Eq. (\ref{eqGMR03}).
\end{widetext}
As expected the PDF of the fraction of occupation time, for the biased CTRW,
depends on the location of the site under consideration. As-usual if $q<1/2$
the particle prefers to stick to the right wall. In our case this behavior
implies that if $q<1/2$ and $x \simeq -L $ ($L$ is large) then $\mathcal{R}%
_x \to 0$, which means that the lattice point $x$ is never occupied, as
expected.

\subsection{Equilibrium--Ergodicity Breaking Relationship}

Eqs. (\ref{eqGMR00}, \ref{eqGMRo2}, \ref{eqGMR03} ) describe the non-ergodic
properties of the CTRW for biased and unbiased cases. We will now consider a
relation of the problem of non-ergodicity with the equilibrium of the
process. Consider an ensemble of independent random walkers performing the
CTRW process, in the finite domain. After a long period of time an
equilibrium will be reached, for which the density of particles is found in
a steady state profile. Such an equilibrium is obtained after each
individual member of the ensemble made many jump events (one can easily
prove that such an equilibrium is reached). We denote the probability of
finding such a random walker on point $x$ with $P_{x} ^{eq}$. It is
straightforward to obtain $P_{x} ^{eq}$, though some care must be made when
we take into consideration the boundary conditions of the problem. In
equilibrium 
\begin{equation}
P_{x} ^{eq} = {\frac{\left( {\frac{1 - q}{q}} \right)^x}{Z}}
\end{equation}
and on the boundaries 
\[
P_{L} ^{eq} = {\frac{\left( 1 - q \right) \left( {\frac{1 - q}{q}}
\right)^{L-1}}{Z}} 
\]
\begin{equation}
P_{-L}^{eq} = {\frac{ q \left( {\frac{1 - q}{q}} \right)^{-L+1}}{Z}}.
\end{equation}
And $Z$ is then obtained from $\sum_{x=-L} ^L P_{x} ^{eq} = 1$. Here $Z$ is
a normalization constant of the problem, not necessarily related to
Boltzmann Gibbs statistics.

Using the equilibrium properties of the system, after a short calculation of
the normalization constant and some algebra, we find that Eqs. (\ref{eqGMR00}%
, \ref{eqGMRo2}, \ref{eqGMR03} ) may be written in a more elegant form 
\begin{equation}
f \left( \overline{p}_{x} \right) =\delta_{ \alpha} \left( {\frac{P_{x}
^{eq} }{1 - P_{x} ^{eq} }},\overline{p}_{x} \right).  \label{eqGMR05aa}
\end{equation}
Note that $P_{x} ^{eq}$ yields the equilibrium properties of many
non-interacting random walkers, or the density profile of large number of
particles. Hence the single particle non-ergodicity is related to
statistical properties of the equilibrium of many particles. The fact that
we find such a relation should be anticipated, since if we average $%
\overline{p}_{x}$ namely consider $\langle \overline{p}_{x} \rangle = \int_0
^\infty \overline{p}_{x} f\left( \overline{p}_{x} \right) \mathrm{d} 
\overline{p}_x$ we must obtain $P_x ^{eq}$, hence $f\left( \overline{p}_{x}
\right)$ must be clearly related to $P_{x} ^{eq}$. And the requirement $%
\langle \overline{p}_{x} \rangle=P_{x} ^{eq}$ implies that $\mathcal{R}_x=
P_{x} ^{eq}/ ( 1 - P_{x} ^{eq})$ as we indeed found (and similar to our
discussion in Sec. \ref{SecLamp} ). For the unbiased case, $q=1/2$ we have $%
P_{x} ^{eq}=1/2 L$, which leads to (\ref{eqGMR00}). Note that the
equilibrium population on the boundaries $x=\pm L$ is half the value of that
found on $x\ne \pm L$, and hence $Z=2 L$ even though we have $2L +1$ lattice
points.

A possible extension of our result: we believe that if we consider the
occupation times on $M<2L+1$ lattice points, Eq. (\ref{eqGen}) is still
valid and $P_{x}^{B}=M/(2L)$ when $L$ is large. A proof of (\ref{eqGen})
based on the calculation of the first passage time for such a case is
cumbersome, if we consider a general configuration of $M$ lattice points
under observations, however we did verify this result numerically.
\begin{figure}[tbp]
\begin{center}
\epsfxsize=70mm \epsfbox{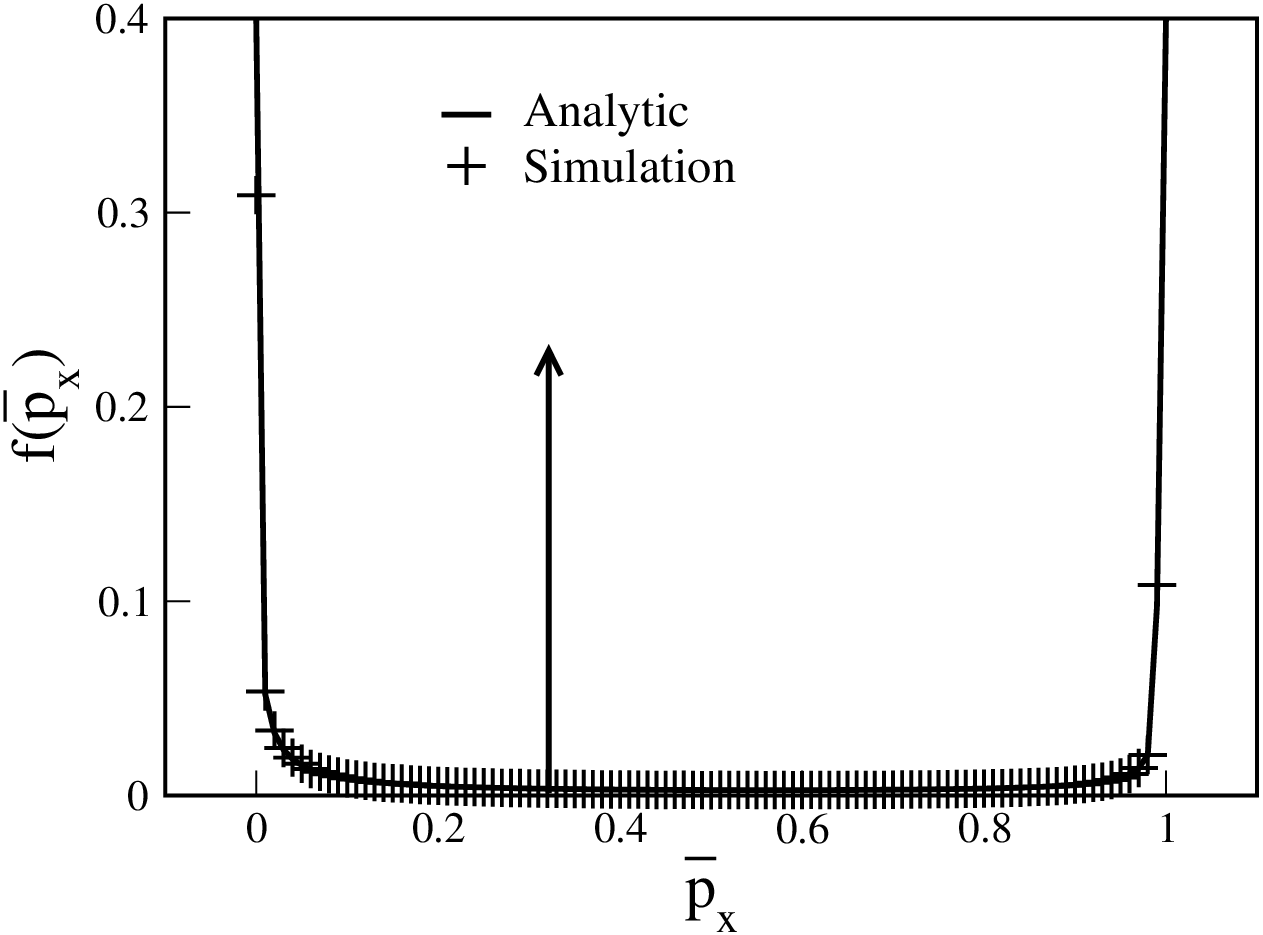}
\end{center}
\caption{ The PDF of occupation times $\overline{p}_x = t_x/t$ where $t_{x}$
is the total time spent on lattice point $x=0$, the minimum of the harmonic
potential field, and $t$ is the measurement time. Here we use $\protect\alpha%
=0.3$. For an ergodic process satisfying detailed balance, the PDF $f(%
\overline{p}_x)$ would be delta centered around the value predicted by
Boltzmann which is given by the arrow. In a given numerical experiment, it
is unlikely to obtain the value of $\overline{p}_x$ predicted by Boltzmann,
though Boltzmann statistics does yield the average of $\overline{p}_x$ over
many measurements. The PDF has a $U$ shape indicating that events, where the
particles hardly ever occupies $x=0$ or nearly always occupies $x=0$ are
important. To construct histograms we used $10^6$ trajectories, measurement
time $t=10^6$, and temperature $T=3$. The solid curve is the analytical
formula Eqs. {\protect\ref{eqGen} \protect\ref{eqGen1}} used without any
fitting parameters. }
\label{fig2}
\end{figure}

\begin{figure}[tbp]
\begin{center}
\epsfxsize=70mm \epsfbox{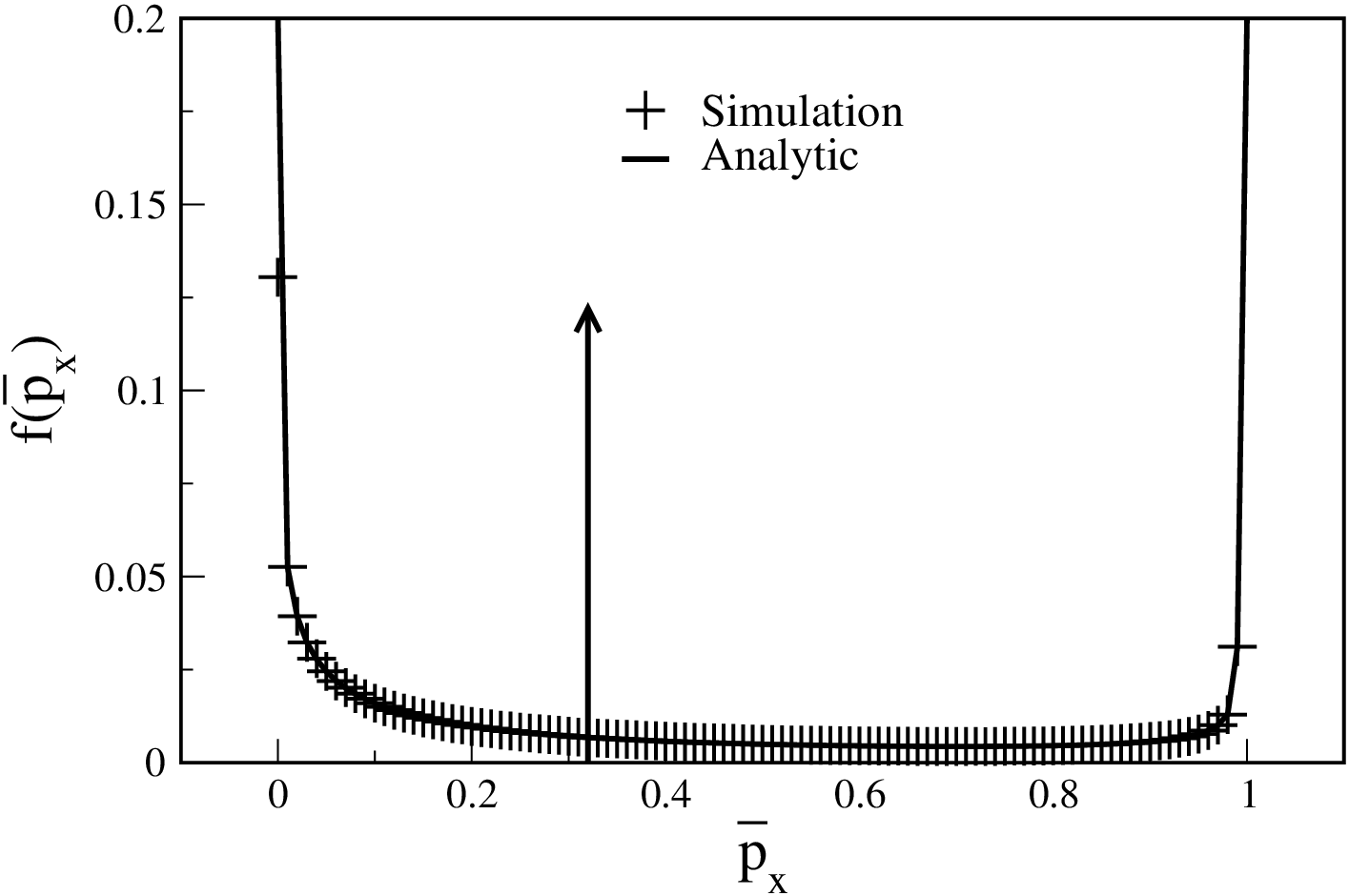}
\end{center}
\caption{ Same as Fig. (\protect\ref{fig2}) however now $\protect\alpha=0.5
$ }
\label{fig3}
\end{figure}

\subsection{Thermal Random Walks}

If the CTRW particle is interacting with a thermal heat bath, we can relate
the non-ergodicity to Boltzmann--Gibbs statistics. For the free particle we
recall that Boltzmann probability of occupying a lattice point is simply 
\begin{equation}
P_{x}^{B}={\frac{1}{Z}}  \label{eqGMR04}
\end{equation}%
and as mentioned $Z=2L$ is the normalization condition, or the partition
function of the problem. Hence rewriting Eq. (\ref{eqGMR05aa}) 
\begin{equation}
f\left( \overline{p}_{x}\right) =\delta _{\alpha }\left( {\frac{P_{x}^{B}}{%
1-P_{x}^{B}}},\overline{p}_{x}\right) .  \label{eqGMR05}
\end{equation}%
The factor $P_{x}^{B}/(1-P_{x}^{B})$ means that with probability $P_{x}^{B}$
the particle is in state $x$, and with probability $1-P_{x}^{B}$ the
particle is in state $nx$ i.e., the rest of the system (here we mean
probability in the ensemble sense).

For biased CTRW when detailed balance condition Eq. (\ref{Eqq}) holds, we
find once again 
\begin{equation}
f\left( \overline{p}_{x}\right) =\delta _{\alpha }\left( {\frac{P_{x}^{B}}{%
1-P_{x}^{B}}},\overline{p}_{x}\right) .  \label{eqGMR06}
\end{equation}%
and now 
\begin{equation}
P_{x}^{B}={\frac{\exp \left( -{\frac{V(x)}{T}}\right) }{Z}},  \label{eqGMR07}
\end{equation}%
where $x$ is the lattice site under observation, $V(x)=-\mathcal{F}ax$ is
the potential field, and $a$ is the lattice spacing. Here the partition
function is 
\begin{equation}
Z=\frac{2\left[ q^{2}\left( {\frac{1-q}{q}}\right) ^{-L+1}-\left( 1-q\right)
\left( {\frac{1-q}{q}}\right) ^{L-1}\right] }{2q-1}.
\end{equation}%
which is easily verified once proper reflecting boundary conditions are
applied, and using Eq. (\ref{Eqq}).

\subsection{Numerical Demonstration}

In previous sections we considered the cases of biased and unbiased CTRWs.
We see however that our results may be more general, and valid also for
random walks in a general deterministic external field. We decided to check
this issue using the example of a random walk in an Harmonic trap. For that
aim we used numerical simulations, since calculations of the first passage
time are cumbersome. The problem of anomalous diffusion in Harmonic
potential was considered in the context of fractional Fokker--Planck
equations \cite{MBK} and in single particle experiments \cite{Xie}.
Anomalous diffusion in harmonic field was also investigated using fractional
Langevin equations \cite{Kou,Kop}. It would be interesting to test if such
stochastic equations yield an ergodic behavior.

The potential field we choose is $V(x)=Kx^{2}$, and $K=1$. We used: (i) the
condition of detailed balance Eq. (\ref{eqDB1}), and (ii) at bottom of the
well, point $x=0$, we used the symmetry of the potential and choose $%
Q_{L}(0)=Q_{R}(0)=1/2$. These two conditions yield $Q_{L}(x)$. In
simulations we generate random waiting times, according to the normalized
power law waiting time PDF $\psi (\tau )=\alpha \tau ^{-(1+\alpha )}$, for $%
\tau >1$.
\begin{figure}[tbp]
\begin{center}
\epsfxsize=70mm \epsfbox{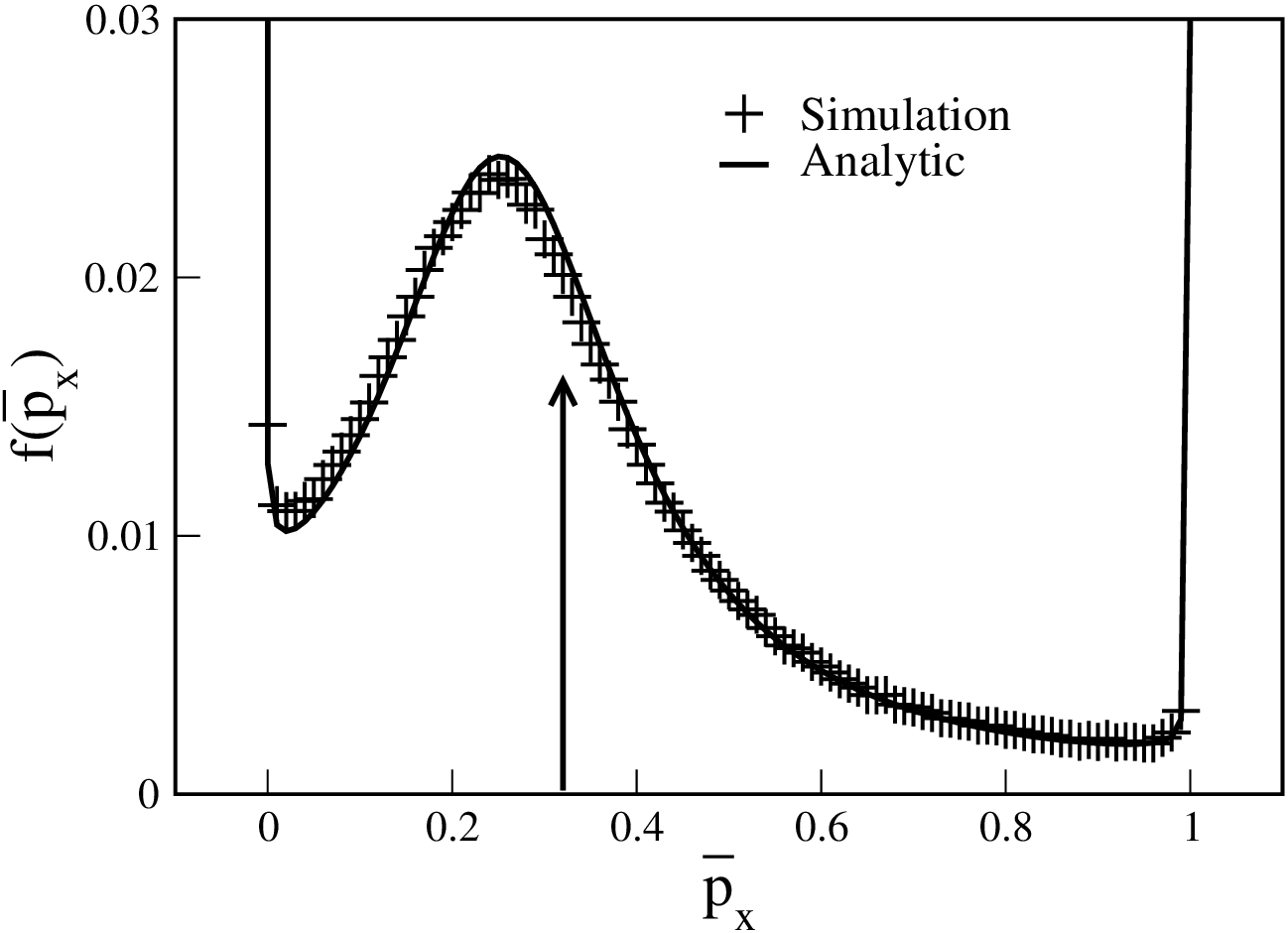}
\end{center}
\caption{ Same as Fig. (\protect\ref{fig2}) however now $\protect\alpha=0.8
$. Unlike Fig. (\protect\ref{fig2}) the PDF has a (distorted) $W$ shape. A
peak close to Boltzmann's value (the arrow) is an indication that as $%
\protect\alpha$ is increased the ergodic phase is approached }
\label{fig4}
\end{figure}

We first checked that Boltzmann equilibrium is reached for an ensemble of
particles. In these simulations we build histograms of the position of $%
N=10^{6}$ particles, after each particle evolves for a time $t=10^{6}$. In
Fig. \ref{fig1} we find good agreement between our simulations and
Boltzmann statistics when many particles are considered. The Fig.
illustrates that an observer of a large number of particles cannot detect
ergodicity breaking, and the single particle limit is essential for our
discussion.

We then consider one trajectory at a time. We obtain from the simulations,
the total time spent by the particle on lattice point $x=0$, namely at the
minimum of the potential. This time is $t_{x}$ and the fraction of
occupation time $\overline{p}_{x}=t_{x}/t$. In the ergodic phase and long
time limit $\overline{p}_{x}$ will approach the value predicted by Boltzmann
statistics. While in the non-ergodic phase we test if our prediction Eq. (%
\ref{eqGen},\ref{eqGen1}) hold.
\begin{figure}[tbp]
\begin{center}
\epsfxsize=70mm \epsfbox{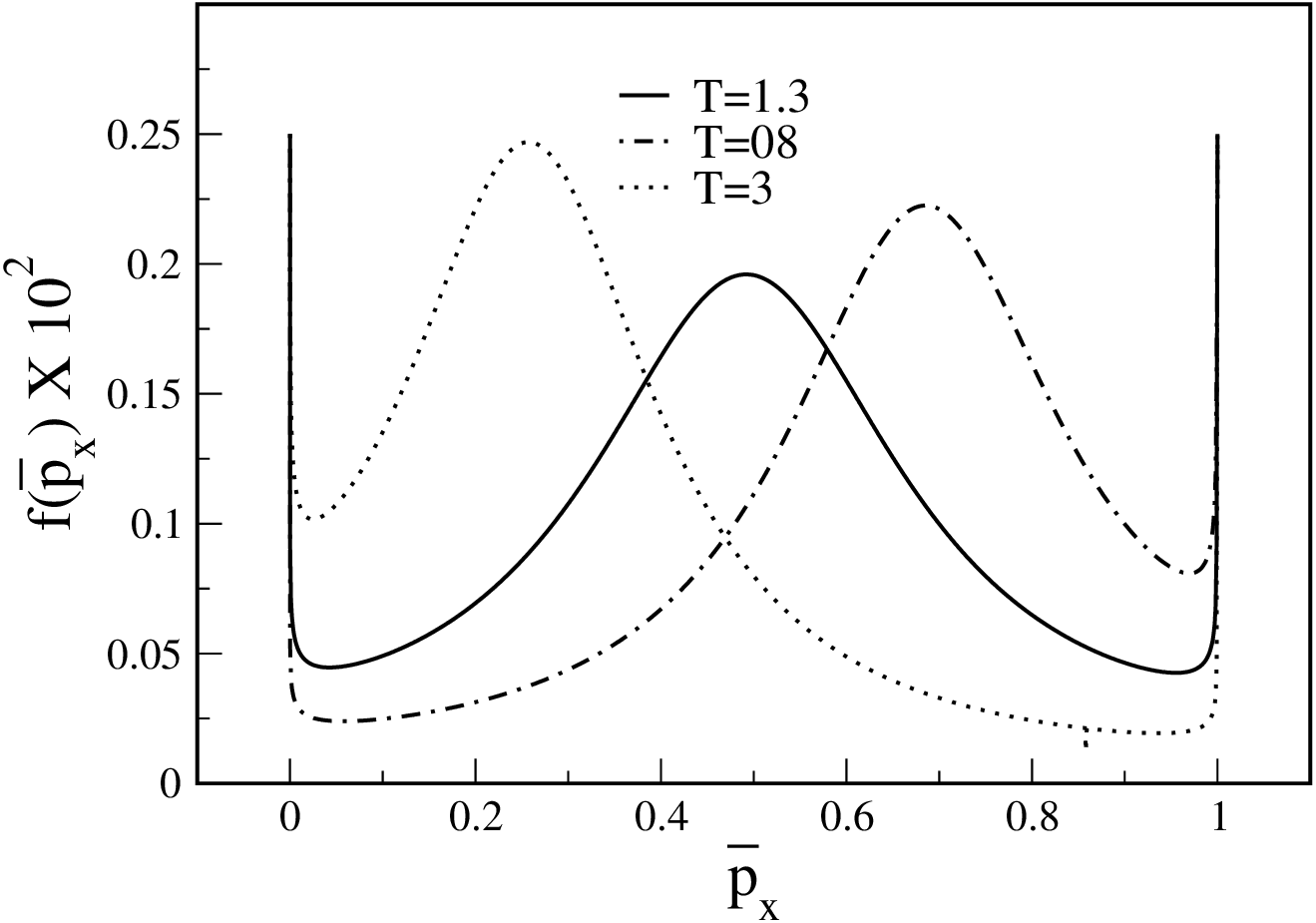}
\end{center}
\caption{ The PDF of fraction of occupation time, for $\protect\alpha=0.8$
when temperature $T$ is varied. Similar to Figs. \protect\ref{fig2}- 
\protect\ref{fig4} we consider occupation times on point $x=0$ of the
Harmonic potential. When temperature is such that probability of occupying $%
x=0$ is equal $1/2$ the PDF is symmetric, for our parameters this temperate
is $T_s=1.3$. For $T << T_s$ the particle tends to be located all the time
on $x=0$, hence for $T=0.8 <T_s$ (the dot-dashed curve) the PDF has more
weight on values $\overline{p}_{x}>1/2$. For $T>>T_s$ the particle is never
found on $x=0$, hence for $T=3>T_s$ (dotted curve) the PDF has more weight
on values of $\overline{p}_{x} < 1/2$. }
\label{fig5}
\end{figure}
In Figs. \ref{fig2}, \ref{fig3}, \ref{fig4} we consider three values of $%
\alpha $, $\alpha =0.3,0.5,0.8$ and fix the temperature $T$. All figures
show an excellent agreement between our theoretical predictions Eqs. (\ref%
{eqGen},\ref{eqGen1}) and numerical simulations. It is more important
however to understand the meaning of the figures.

For small $0<\alpha <<1$ we expect that the particle will get stuck on one
lattice point during a very long period, which is of the order of the
measurement time $t$. This trapping point, can be either the point of
observation (e.g. $x=0$ in our simulations) or some other lattice point. In
these cases we expect to find $\overline{p}_{x}\simeq 1$ or $\overline{p}%
_{x}\simeq 0$ respectively. Hence the PDF of $\overline{p}_{x}$ has a $U$
shape. This case exhibits large deviations from ergodic behavior, in the
sense that we have a very small probability for finding the occupation
fraction close to the value predicted based on Boltzmann's ergodic theory.
As shown in Figs. (\ref{fig2}, \ref{fig3}) such $U$ shape behavior is found
for the cases $\alpha =0.3$ and $\alpha =0.5$. We also plotted the
prediction made using the ergodic assumption (the arrows in the Figures) to
demonstrate the fact that a measurement is not likely to yield the average
which is located on $P_{x}^{B}$.

When we increase $\alpha $ we anticipate a more ergodic behavior, in
particular in the limit $\alpha \rightarrow 1$. An ergodic behavior means
that the PDF of the occupation fraction $\overline{p}_{x}$ is centered on
the Boltzmann's probability. In Fig. \ref{fig4} where $\alpha =0.8$ we start
seeing a peak in the PDF of $\overline{p}_{x}$ centered in the vicinity of
the ensemble average value. Note however that the PDF $f(\overline{p}_{x})$
still attains its maxima on $\overline{p}_{x}=0$ and $\overline{p}_{x}=1$.
Hence we find a weaker non-ergodic behavior if compared with the cases $%
\alpha =0.3,0.5$, and a $W$ shape of the PDF.

In Fig. \ref{fig5} we consider the example of a particle in an Harmonic
field, we fix $\alpha =0.8$ and vary temperature, using Eqs. (\ref{eqGen},%
\ref{eqGen1}). The observation point remains $x=0$. At temperature $T\simeq
1.3$ (solid line) we see that the PDF of $\overline{p}_{x}$ is symmetric.
This happens when $P_{x}^{B}=1/2$, namely for a case that there is
probability half of occupying the observation point, and probability half to
be out of this point. When the temperature is very low, we expect to find
the particle, in the ground state, namely on $x=0$. Hence the PDF of $%
\overline{p}_{x}$ is tilted towards $\overline{p}_{x}\simeq 1$ when
temperature is lowered (see Fig.\ref{fig5} when $T=0.8$ ). In contrast when
the temperature is high, we expect the probability of occupying the
observation point $x=0$ to be reduced (as-usual entropy wins at high
temperature). And indeed we observe that when $T=3$ the PDF of $\overline{p}%
_{x}$ is more tilted towards the left namely to $\overline{p}_{x}\simeq 0$.

\subsection{Validity of main Eqs. (\protect\ref{eqGen},\protect\ref{eqGen1})}

Our numerical work as well as our analytical solutions for the bias and
non-bias CTRW show the validity of Eqs. (\ref{eqGen},\ref{eqGen1}). What
happens for more general type of potential fields? Can we claim, that Eq. (%
\ref{eqGen}) has a wider applicability? Consider the CTRW with potential
profile, $\{ ... E_x ...\}$, with the dynamics satisfying detailed balance
condition. We claim, but have no rigorous proof, that if for $\psi(\tau)$
with finite moments, the system is ergodic, then for the same energy profile
but when the waiting time has a long tail, Eqs. (\ref{eqGen},\ref{eqGen1})
hold. Our reasoning is that we can think of $\alpha$ as a control parameter,
which we can vary between $0<\alpha\le 1$. And since for the case $\alpha=1$
we have $\mathcal{R}_x = P_x ^B/ (1 - P_x ^B)$ also for $0<\alpha<1$ this
relation must hold (since $P_B$ does not depend on $\alpha$). Further the
transformation $s\langle \tau \rangle \to A s^{\alpha}$ in the small $s$
behavior of the waiting time seems to indicate that the behavior we found
has a general validity.

A way to understand the ergodicity breaking laws, Eqs. (\ref{eqGen},\ref%
{eqGen1}), is to consider the number of times $n_x$ the particle visits
lattice point $x$, during a long measurement time. In that case the particle
visits $x$ many times, and we assume that the fraction of number of visits
satisfies 
\begin{equation}
{\frac{ n_x }{n }} = \exp\left[ - {\frac{V(x) }{T}} \right],  \label{Eqrat8}
\end{equation}
where $n$ is the total number of jumps made by the particle. If the first
moment of the waiting time distribution is finite, we have $t_x /t = n_x /n$%
, since the average time spent on $x$ is $n_x$ times the mean waiting time.
When the average waiting time is infinite $\alpha<1$, one can show that the
PDF of $t_x /t$ is given by Eq. (\ref{eqGen}) if condition Eq. (\ref{Eqrat8}%
) holds. Eq. (\ref{Eqrat8}) should be tested in more detail, for example
using numerical simulations.

\section{Summary and Discussion}

We obtained the non ergodic properties of biased and unbiased continuous
time random walks. In particular the distribution of the occupation fraction 
$\overline{p}_{x}$ was found. Our results are valid both for thermal and
non-thermal cases. In both cases the non-ergodicity is described using the $%
\delta _{\alpha }(\mathcal{R}_{x},\overline{p}_{x})$ PDF Eq. (\ref{eqGen}).
Where $\alpha $ is the anomalous diffusion exponent $\langle x^{2}\rangle
\sim t^{\alpha }$. Both for thermal and non-thermal random walks the
parameter $\mathcal{R}_{x}$ is related to the ensemble averaged equilibrium
properties of the system, Eqs. (\ref{eqGMR05aa}, \ref{eqGMR06})
respectively. If the system is in vicinity of thermal equilibrium, the
equilibrium of the system is the Boltzmann--Gibbs equilibrium, in the
ensemble sense. Such behavior is found when detailed balance conditions are
applied. In this case the characterization of the non-ergodic properties of
the occupation times is related to the partition function, and temperature.
The non-ergodicity manifests itself when time average of single particle
observables is considered. In particular the occupation time, in a given
energy state, or on a particular lattice point. Hence the non-ergodicity
might reveal itself in single particle experiments.

Models and systems describing anomalous diffusion are wide spread. In most
cases ensemble average properties of such processes are investigated, both
in theory and in experiment. For single particle experiments, where the
problem of ensemble averaging is removed, we may either: i) reconstruct the
ensemble averages, by repeating the single molecule experiment many times,
or ii) investigate the ergodic properties of the system, by considering
fraction of occupation time in a particular state, and obtaining its
distribution. It is the second type of measurement which is considered here,
which yields insight into single particle properties which differ from the
standard ensemble measurement, provided that a non-ergodic phase is
investigated. And while the theory of anomalous diffusion processes is now
vast, the non-ergodic properties of such processes are still not well
understood. Investigation of this topic beyond the CTRW approach is left for
future work.

\textbf{Acknowledgment:} EB was supported by the National Science Foundation
award No. CHE-0344930, and the Center for Complexity Science Jerusalem. EB
thanks J. P. Bouchaud, A. Comtet, S. N. Majumdar, and G. Margolin for
comments and discussion.

\section{Appendix A: The $\protect\delta_{\protect\alpha}\left( \mathcal{R}%
_x, \overline{p}_{x} \right)$ Function}

\label{SecARC}

In this Appendix we re-derive the limit theorem Eq. (\ref{eqGen}). While
this goal was accomplished long time ago \cite{Lamp}, we believe that it is
worth while re-deriving this result using methods similar to those used
today in statistical Physics community \cite{Godreche}. We also derive an
exact distribution for the occupation fraction of a two state process in
Laplace space Eq. (\ref{eqGA08}, \ref{eqGA09}).
\begin{figure}[tbp]
\begin{center}
\epsfxsize=70mm \epsfbox{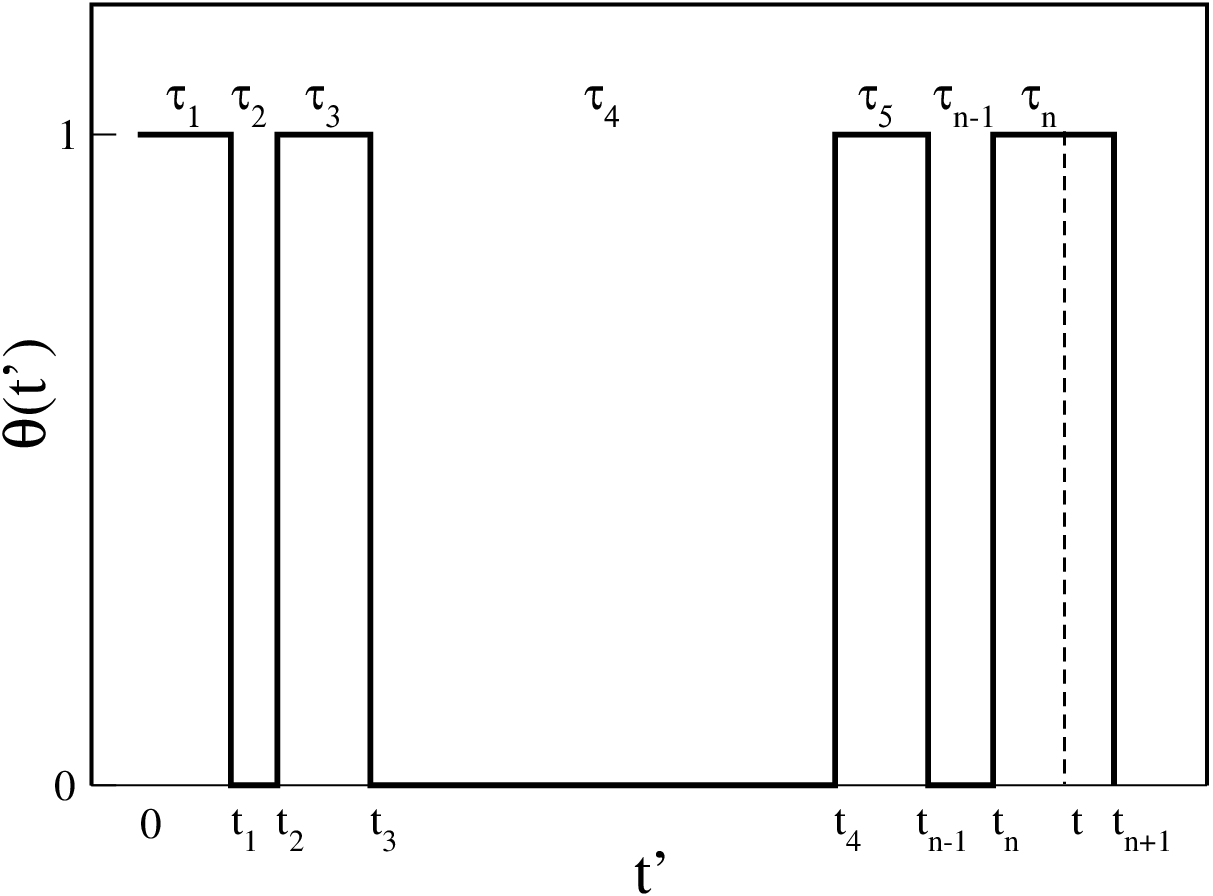}
\end{center}
\caption{ A schematic diagram of the two state process. }
\label{fig6}
\end{figure}

Consider a system evolving between two states, $+$ and $-$, corresponding to
states $x$ and $nx$ respectively. Let $\theta (t)=1$ when the system is in
state $+$ otherwise $\theta (t)=0$ and the system is in state $-$. A
schematic diagram of $\theta (t)$ is shown in Fig. \ref{fig6}. Let $%
\{t_{i}\} $ denote dots on the time axis on which transition events between
state $+$ to state $-$ or vice versa is made. Let $\{\tau _{i}\}$ be sojourn
times either in state $+$ or state $-$. If the process starts with state $+$%
, then $\tau _{i}$ is a $+$ state if $i$ is odd. We also denote the total
number of jumps, in the measurement time interval $(0,t)$ with $n$. We
assume that the sojourn times are independent identically distributed random
variables. The PDF of sojourn times is $\psi _{+}(\tau )$ and $\psi
_{-}(\tau )$ for states $+$ and $-$ respectively. Such a simple process is
called a two state renewal process. As-usual it is convenient to analyze
such a stochastic process using the Laplace transforms 
\begin{equation}
\hat{\psi}_{\pm }\left( s\right) =\int_{0}^{\infty }e^{-s\tau }\psi _{\pm
}\left( \tau \right) \mathrm{d}\tau .  \label{eqGA05}
\end{equation}

While we will consider general properties of the stochastic process, we will
eventually focus on two main cases. First consider the case where all
moments of $\psi_{\pm}(\tau)$ are finite, e.g. exponential PDFs belong to
this category. Then the following small $s$ expansion holds 
\begin{equation}
\hat{\psi}_{\pm} (s) \sim 1 - s \langle \tau_{\pm} \rangle + \cdots.
\label{eqGAaa}
\end{equation}
Here $\langle \tau_{\pm}\rangle$ are the mean sojourn times in states $\pm$.
A second generic case is: 
\begin{equation}
\hat{\psi}_{\pm} (s) \sim 1 - s^{\alpha} A_{\pm} + \cdots.  \label{eqGAaa1}
\end{equation}
with $0<\alpha < 1$. For example the one sided L\'evy PDF $\hat{\psi}_{\pm}
(s) = \exp( - A_{\pm} s^{\alpha} )$ belongs to this class. In the time
domain these PDFs behave like 
\begin{equation}
\psi_{\pm} (t) \sim {\frac{ A_{\pm} }{|\Gamma\left(-\alpha \right)| t^{ 1 +
\alpha} }}  \label{eqLTsss}
\end{equation}
when $t$ is large, namely for this family of PDFs the average waiting times
in both states diverges.

Let $t_{+}$ be the total time spent in state $+$, within the time period $%
(0,t)$. Then the occupation fraction in state $+$ is 
\begin{equation}
\overline{p}_{+}={\frac{t_{+}}{t}}={\frac{\int_{0}^{t}\theta (t^{\prime })%
\mathrm{d}t^{\prime }}{t}}.  \label{eqTAzz}
\end{equation}%
We now consider statistical properties of $t_{+}$ focusing on the scaling
limit $t\rightarrow \infty $. Let $f_{t,n}^{+}\left( t_{+}\right) $ be the
PDF of $t_{+}$ conditioned that $n$ renewal (i.e. jumps) events occurred in
the time interval $(0,t)$, and that the start of the process is in state $+$.

Consider the case $n$ odd, $n=2k+1$ with $k=0,1,\cdots $. Then since we
start with state $+$, $t_{+}=\sum_{i=1,\mbox{odd}}^{n}\tau _{i}$ where the
summation is only over odd $i$'s. Also we have $t_{n}<t<t_{n+1}$, where $%
t_{n+1}$ is a renewal event which occurs after end of measurement (see Fig. %
\ref{fig6}). Hence 
\begin{equation}
f_{t,n}^{+}\left( t_{+}\right) =\langle \delta \left( t_{+}-\sum_{i=1,%
\mbox{odd}}^{n}\tau _{i}\right) I\left( t_{n}\leq t\leq t_{n+1}\right)
\rangle ,  \label{eqGA01}
\end{equation}%
where $\delta (x)$ is the Dirac delta function, and 
\begin{equation}
I\left( t_{n}\leq t\leq t_{n+1}\right) =\left\{ 
\begin{array}{ll}
1 & \mbox{ If condition in parenthesis is true} \\ 
0 & \mbox{otherwise}.%
\end{array}%
\right.   \label{eqGA02}
\end{equation}%
In Eq. (\ref{eqGA01}) $\langle \cdots \rangle $ denotes an average over the
stochastic process, soon to be specified in more detail. Later we will
consider the case $n$ even, and then sum over $n$ to obtain the PDF of $t_{+}
$.

Let $s$ be the Laplace pair of $t$, and $u$ of $t_{+}$. It is convenient to
consider the double Laplace transform $\hat{f}_{n,s}^{+}\left( u\right) $ of 
$f_{n,t}^{+}\left( t_{+}\right) $ 
\[
\hat{f}_{n,s}^{+}\left( u\right) =\int_{0}^{\infty
}e^{-t_{+}u}\int_{0}^{\infty }e^{-st}f_{n,t}^{+}\left( t_{+}\right) \mathrm{d%
}t_{+}\mathrm{d}t=
\]%
\[
\langle \int_{0}^{\infty }\int_{0}^{\infty }e^{-ut_{+}-st}\delta \left(
t_{+}-\sum_{i=1,\mbox{odd}}^{n}\tau _{i}\right) I\left( t_{n}\leq t\leq
t_{n+1}\right) \mathrm{d}t_{+}\mathrm{d}t\rangle 
\]%
$\ \ \ $ 
\begin{equation}
=\langle {\frac{e^{-st_{n}}-e^{-st_{n+1}}}{s}}e^{-u\sum_{i=1,\mbox{odd}%
}^{n}\tau _{i}}\rangle ,  \label{eqGA03}
\end{equation}%
where we made use of Eq. (\ref{eqGA01}). Now we may consider the average $%
\langle \cdots \rangle $, using $t_{n}=\sum_{i=1,\mbox{odd}}^{n}\tau
_{i}+\sum_{i=2,\mbox{even}}^{n-1}\tau _{i}$ and similarly for $t_{n+1}$.
Recalling that $\tau _{i}$ with odd (even) $i$ are $+$ $(-)$ states
respectively we find after averaging over the $\{\tau _{i}\}$s 
\begin{equation}
\hat{f}_{n,s}^{+}\left( u\right) =\hat{\psi}_{+}^{k+1}\left( s+u\right) \hat{%
\psi}_{-}^{k}\left( s\right) {\frac{1-\hat{\psi}_{-}\left( s\right) }{s}}
\label{eqGA04}
\end{equation}%
where $n=2k+1$. For even $n$ such that $n=2k$, $k=1,2,\cdots $ we obtain 
\begin{equation}
\hat{f}_{n,s}^{+}\left( u\right) =\hat{\psi}_{+}^{k}\left( s+u\right) \hat{%
\psi}_{-}^{k}\left( s\right) {\frac{1-\hat{\psi}_{+}\left( s+u\right) }{s+u}}%
.  \label{eqGA06}
\end{equation}%
Note that for even $n$ the last $+$ interval falls on $t$, and we must
define $\tau _{n}=t-t_{n}$ as the time difference between end of the
measurement and last jump in the sequence (see Fig. \ref{fig6}). We are
ready to obtain the double Laplace transform of $f_{t}^{+}(t_{+})$, i.e.,
the PDF of $t_{+}$ when the process starts in $+$ state, 
\begin{equation}
\hat{f}_{s}^{+}\left( u\right) =\sum_{k=0}^{\infty }\left[ \hat{f}%
_{2k,s}^{+}\left( u\right) +\hat{f}_{2k+1,s}^{+}\left( u\right) \right] .
\label{eqGA07}
\end{equation}%
Using Eqs. (\ref{eqGA04},\ref{eqGA06},\ref{eqGA07}) we obtain the exact
solution to the problem in Laplace $s$, $u$ space 
\begin{widetext}
\begin{equation}
\hat{f}^{+}_{s} \left( u \right) = \left[ { 1 - \hat{\psi}_{+} \left( s + u \right) \over
s + u } + \hat{\psi}_{+} \left( s + u \right) { 1 - \hat{\psi}_{-} \left( s \right) \over s} \right]
{ 1 \over 1 - \hat{\psi}_{+} \left( s + u \right) \hat{\psi}_{-} \left( s \right) }.
\label{eqGA08}
\end{equation}
It is easy to check the normalization condition $f^{+}_s(u=0)=1/s$ provided
of-course that $\psi_{\pm}(\tau)$ are normalized PDFs. 
In a similar way one can show that if we start the process
in state $-$ the double Laplace transform
of the PDF of $t_{+}$ denoted with 
$f_{t} ^{-} \left( t_{+} \right)$ is
\begin{equation}
\hat{f}^{-}_{s} \left( u \right) = 
\left[\hat{\psi}_{-}\left( s \right)  { 1 - \hat{\psi}_{+} \left( s + u \right) \over
s + u } +{1- \hat{\psi}_{-} \left( s\right) \over s} \right]
{ 1 \over 1 - \hat{\psi}_{+} \left( s + u \right) \hat{\psi}_{-} \left( s \right) }.
\label{eqGA09}
\end{equation}
\end{widetext}Eqs. (\ref{eqGA08}, \ref{eqGA09}) yield in principle the exact
expression for the occupation fraction, which might be useful in determining
the pre-asymptotic behavior, for example using numerical inverse Laplace
transform.

For the generic case Eq. (\ref{eqGAaa1}), in the limit of $s \to 0$ and $u
\to 0$ their ratio remaining arbitrary, Eqs. (\ref{eqGA08},\ref{eqGA09})
yield 
\begin{equation}
\hat{f}^{\pm} _s \left( u \right) \sim {\frac{ \mathcal{R} \left( s + u
\right)^{\alpha -1} + s^{\alpha -1} }{\mathcal{R} \left( s + u
\right)^{\alpha} + s^{\alpha} }}  \label{eqGA10}
\end{equation}
with 
\begin{equation}
\mathcal{R} = {\frac{ A_{+}}{A_{-}}}.  \label{eqAm25}
\end{equation}
The amplitude ratio $\mathcal{R}$ determines the degree of symmetry in the
problem. Note that in this scaling limit the initial state of the process,
i.e. process being in state $+$ or $-$ at initial time, is not important.

The small $(s,u)$ limit considered in Eq. (\ref{eqGA10}) corresponds to
large measurement time $t$, and occupation time $t_{+}$ limit. We invert Eq.
(\ref{eqGA10}) using a method given in \cite{Godreche}. The method states
that if in the limit $s,u\rightarrow 0$ a double Laplace transform behaves
like 
\begin{equation}
\hat{f}_{s}\left( u\right) ={\frac{1}{s}}g\left( {\frac{s}{u}}\right) 
\end{equation}%
then the PDF of the scaled variable $\overline{p}_{+}=t_{+}/t$ is in the
long time $t$ limit 
\begin{equation}
f\left( \overline{p}_{+}\right) =-{\frac{1}{\pi x}}\lim_{\epsilon
\rightarrow 0}\mbox{Im}g\left( -{\frac{1}{x+i\epsilon }}\right) |_{x=%
\overline{p}_{+}}
\end{equation}%
Using Eq. (\ref{eqGA10}) we find the PDF of the fraction of occupation time $%
\overline{p}_{+}=t_{+}/t$ 
\[
f\left( \overline{p}_{+}\right) =\delta _{\alpha }\left( \mathcal{R},%
\overline{p}_{+}\right) =
\]%
\begin{equation}
{\frac{\sin \pi \alpha }{\pi }}{\frac{\mathcal{R}\overline{p}_{+}^{\alpha
-1}\left( 1-\overline{p}_{+}\right) ^{\alpha -1}}{\mathcal{R}^{2}\left( 1-%
\overline{p}_{+}\right) ^{2\alpha }+\overline{p}_{+}^{2\alpha }+2\mathcal{R}%
\left( 1-\overline{p}_{+}\right) ^{\alpha }\overline{p}_{+}^{\alpha }\cos
\pi \alpha }}  \label{eqGA11}
\end{equation}%
The PDF is normalized according to $\int_{0}^{1}f(\overline{p}_{+})\mathrm{d}%
\overline{p}_{+}=1$, it is valid only in the long time $t$ limit and is
independent of it. In this sense an equilibrium is obtained. In particular
when $A_{+}=A_{-}$ and $\alpha =1/2$ we find the arcsin distribution. It is
easy to show that the average 
\begin{equation}
\langle \overline{p}_{+}\rangle ={\frac{\langle t_{+}\rangle }{t}}={\frac{%
A_{+}}{A_{+}+A_{-}}}.
\end{equation}%
In the limit $\alpha \rightarrow 1$ we obtain 
\begin{equation}
f\left( \overline{p}_{+}\right) =\delta \left( \overline{p}_{+}-{\frac{%
\langle \tau \rangle _{+}}{\langle \tau \rangle _{+}+\langle \tau \rangle
_{-}}}\right) ,
\end{equation}%
where $\langle \tau \rangle _{\pm }$ are the average waiting times when the
waiting time PDFs have finite moments. We identify this behavior with an
ergodic behavior, since according to Eq. (\ref{eqTAzz}) $\overline{p}_{+}$
is a time average of $\theta (t)$, which is equal to the ensemble average
value when moments of $\psi _{\pm }(\tau )$ are finite.

From Eq. (\ref{eqGA11}) we see that the PDF of $\overline{p}_{+}$ is not
narrowly centered on the ensemble average value when $\alpha<1$. Hence the
case $\alpha<1$ is called the non-ergodic phase. Indeed any measurement of $%
\overline{p}_{+}$ is unlikely to yield the average, and $f \left( \overline{p%
}_{+} \right) \to \infty$ when $\overline{p}_{+} \to 0$ or $\overline{p}_{+}
\to 1$. The latter behavior corresponding to systems in state $-$ or state $%
+ $ during nearly all the measurement time, respectively. The reason for
non-ergodic behavior is that in a measurement time interval $(0,t)$ we
expect to obtain a few sojourn times of the order of $t$, these can be
either $+$ times or $-$ times or both. In each measurement we make these
large sojourn times will be different than those found in a second
measurement. Hence the time average of $\theta(t)$, the occupation fraction $%
\overline{p}_{+}$ will remain random even in the long time limit.

\end{document}